\documentclass[nofootinbib,showpacs,preprintnumbers,amsmath,amssymb]{revtex4}

\usepackage{graphicx}
\usepackage{dcolumn}
\usepackage{bm}
\usepackage{hyperref} 
\usepackage{lipsum}  %
\usepackage{color} 
\usepackage[pdftex]{epsfig}

\def\square{\kern1pt\vbox{\hrule height 1.2pt\hbox{\vrule width 1.2pt\hskip 3pt
   \vbox{\vskip 6pt}\hskip 3pt\vrule width 0.6pt}\hrule height 0.6pt}\kern1pt}

\newcommand{\be}{\begin{equation}}
\newcommand{\ee}{\end{equation}}
\newcommand{\bea}{\begin{eqnarray}}
\newcommand{\eea}{\end{eqnarray}}
\newcommand{\nn}{\nonumber}

\begin{document}

\title{Quantum gravity corrections to the conformally coupled scalar self-mass-squared 
on de Sitter background}

\author{Sibel Boran}
\email{borans@itu.edu.tr}
\author{E. O. Kahya}
\email{eokahya@itu.edu.tr}
\affiliation{Department of Physics, \.Istanbul Technical University, Maslak 34469 Istanbul, Turkey}
\author{Sohyun Park}
\email{spark@gravity.psu.edu}
\affiliation{Institute for Gravitation and the Cosmos, The Pennsylvania State University, University
Park, PA 16802, USA}

\date{\today}
\begin{abstract}
We evaluate one loop quantum gravity corrections to the conformally coupled (CC)
scalar self-mass-squared on a locally de Sitter background. 
In this paper we consider only the conformal-conformal interaction part of the self-mass-squared. 
This complements the minimal-minimal part worked out in the previous paper \cite{kahya} and 
we will add the minimal-conformal part in a follow-up paper to complete 
the full self-mass-squared at one loop order. 
The computation is performed using dimensional regularization and  
the results are fully renormalized by absorbing divergences with counterterms.
The finite results can give rise to quantum corrections to the CC scalar mode functions
and therefore to their power spectra. 
\end{abstract}

\maketitle

\section{Introduction}\label{introduction}

An intuitive way of understanding quantum effects is to examine the classical 
response to virtual particles. If there are not many virtual particles or they 
interact with the particles in question only weakly there will not be much 
quantum loop effect. In this respect it will be interesting to investigate 
the cases in which the number of virtual particles increases and their interactions 
with other particles. It has been known for a long time that the expansion of 
spacetime can lead to particle creation by delaying the annihilation of virtual 
pairs ripped out of the vacuum since Schr\"odinger first realized it \cite{Schr}.
In late 1960's Parker carried out explicit computations regarding particle production 
in expanding universe \cite{Parker1} and the results are summarized as follows: 
the effect is the strongest if the expansion is accelerated, and the virtual 
particles are massless and do not possess conformal symmetry \cite{Parker2}.

The locally de Sitter background we consider is the most highly accelerated 
expansion with classical stability. 
The unique examples of massless particles with no conformal invariance
are the massless, minimally coupled (MMC) scalars and gravitons. These particles 
are the sources of the scalar and tensor perturbations predicted by inflationary 
theories \cite{SMC}. The scalar component of which has been mapped with great 
precision \cite{WMAP9, Planck16}. The tensor components have not been mapped yet, 
but with the recent announcement for the detection of the $B$-mode polarization 
by the BICEP2 collaboration \cite{BICEP2}, it is not hopeless to image them in 
the near future. 

The approximate tree order results for the scalar and tensor power spectra are
\be
\Delta^2_{\mathcal{R}}(k) =  \lim_{t \gg t_k}\frac{k^3}{2\pi^2}\times4\pi G \times \vert v(t,k)\vert^2  \simeq \frac{G H^2}{\pi \epsilon}\;, \quad 
\Delta^2_{h}(k) = \lim_{t \gg t_k}\frac{k^3}{2\pi^2}\times64\pi G \times \vert u(t,k)\vert^2 \simeq \frac{16G H^2}{\pi}\;.
\label{powerspec}
\ee
Here $v(t,k)$ and $u(t,k)$ are the mode functions of scalar and tensor perturbations, 
$G$ is Newton's constant, $H(t)$ is the Hubble parameter 
and $\epsilon(t)$ is the slow roll parameter. 
The time $t$ is taken to be much later than $t_k$, the time of first horizon crossing for the mode of wave number $k$. 
One can study loop corrections to this lowest order effect 
using a mathematical object called the one-particle-irreducible (1PI) 2-point function. 
The procedure is first to compute the 1PI 2-point function, for the case of scalar 
it is the self-mass squared, $-iM^2(x;x')$. Second, to use it to quantum-correct 
the linearized effective field equation, for the CC scalar it is
\be
\partial_{\mu}\Bigl(\sqrt{-g}g^{\mu\nu}\partial_{\nu}\phi(x)\Bigr)
-\frac{1}{6}R\phi(x) - \int d^4x' M^2(x;x')\phi(x') = 0\;.
\label{linear_eq}
\ee   
By solving this equation one can get loop corrections to the mode functions for the inflaton, where the conformal coupling term can be thought effectively as an inflaton potential (similar to the $m^2 \phi^2$ model) in the context of scalar-driven inflation models at the lowest order.

A number of computations in this regard have been made over the last decade 
\cite{phi4, SQED, PP, Yukawa, mw, SPM, KW1, kahya, kahya2, PW, LW, PMTW, LPPW,
CV, FRV1, FPRV, FRV2, Frob
DS2, DS1, UM, DS3, 
MM,
ABG, YK, APS, AK, YU, LBH, Akh, GRZ}. 
(A simple worked-out example specifically for loop corrections to the power 
spectra can be found in \cite{MP}.) In this paper we carry out the first part 
of this procedure, that is we  compute the self-mass-squared of CC 
scalar interacting with inflationary gravitons at one loop order and 
renormalize it so to make the integral in (\ref{linear_eq}) finite. In a follow-up 
paper we will solve the effective field equation to study one loop corrections 
to the tree order mode functions of the CC scalar field. 

One might expect strong quantum effects for the interaction between the MMC scalars and gravitons because 
these two particles (being massless and without conformal invariance) are the ones produced 
enormously during inflation \cite{Parker2}. However, many studies \cite{phi4, SQED, 
PP, Yukawa, mw, SPM, KW1, kahya, kahya2, PW, LW, PMTW, LPPW} suggest that the strongest 
effects are caused by non-derivative interactions. The reason why it happens is
the differentiated scalar, so its kinetic energy is redshifted away during inflation 
so they interact very weakly with virtual gravitons. So one loop corrections from gravitons 
to the self-mass-squared of MMC scalar result in zero \cite{kahya, kahya2}. 
The same phenomenon happens to the mirror case, namely the differentiated gravitons 
interacting with virtual scalars \cite{PW,LPPW}. In fact, these null results motivated 
the authors to consider the conformally coupled scalar interacting with virtual 
gravitons by noting that in the conformal coupling term the scalar 
is not differentiated. Moreover, the mode functions of conformally coupled scalar 
redshift to zero at tree order. Therefore any loop corrections to them would be 
dominant over the zero tree order result.

In Section \ref{setup}, we derive a formal expression for the CC scalar self-mass-squared. At one-loop order it consists of two parts, namely 4-point and 3-point interactions.  
Explicit computations of the two parts are performed in Section \ref{ccssm}. 
The results are fully renormalized using appropriate counterterms
in Section \ref{renormalization}. Our discussions comprise Section \ref{dis}.

\section{The self-mass-squared}\label{setup}

The Lagrangian which describes pure gravity and the interaction between gravitons 
and the conformally coupled scalar (in $D$ spacetime dimensions to facilitate dimensional regularization) is
\bea
\label{Lagrangian}
\mathcal{L} = 
- \frac12 \partial_{\mu} \phi \partial_{\nu} \phi g^{\mu\nu} \sqrt{-g} - \frac{D-2}{8(D-1)}\phi^2 R \sqrt{-g} 
+ \frac{1}{16 \pi G} (R-(D-2)\Lambda )\sqrt{-g}\;,
\eea
where $R$ is the Ricci scalar and $\Lambda = (D-1) H^2$ is the cosmological constant.
Varying the 1PI effective action corresponding to the Lagrangian \eqref{Lagrangian}
with respect to the scalar field we have the linearized effective field equation \eqref{linear_eq},
\bea
\partial_{\mu}\Bigl(\sqrt{-g}g^{\mu\nu}\partial_{\nu}\phi(x)\Bigr)
-\frac{1}{6}R\phi(x) - \int d^4x' M^2(x;x')\phi(x') = 0\;.
\nn
\eea   

We work on the open conformal coordinate patch of de Sitter space
\be
ds^2 = \hat{g}_{\mu\nu}dx^{\mu}dx^{\nu} 
= a^2(\eta) \Bigl[ -d\eta^2 + d\vec{x} \!\cdot\! d\vec{x}
\Bigr] 
\;, \quad \mbox{where} \quad a(\eta) = -\frac1{H \eta} \;,
\label{bkgd_geometry}
\ee
with the coordinate ranges
\be
-\infty < x^0 \equiv \eta < 0 \quad , \quad -\infty < x^i <
+\infty \quad, \quad i = 1\; 2\;,\; \cdots , D\!-\!1 \; .
\ee
Perturbation theory is expressed in terms of 
the background metric $\hat{g}_{\mu\nu}$ and conformally rescaled graviton field $h_{\mu\nu}$
\be
g_{\mu\nu}(x) \equiv a^2 \Bigl[\eta_{\mu\nu} + \kappa h_{\mu\nu}(x)\Bigr]\;, 
\quad \mbox{where} \quad \kappa^2 \equiv 16\pi G\;,
\ee
Due to the conformal coupling term (the second term) 
the matter sector (the first and second term) in the Lagrangian \eqref{Lagrangian}
is invariant under the conformal rescaling,  
\bea
g_{\mu\nu} &\equiv&  \Omega^2 \tilde{g}_{\mu\nu}\;, \quad
\phi \equiv  \Omega^{\frac{2-D}{2}} \tilde{\phi}\;.
\label{rescaling} \\
\Rightarrow
\mathcal{L}_{\rm Mat} &=& 
- \frac12 \partial_{\mu} \phi \partial_{\nu} \phi g^{\mu\nu} \sqrt{-g} 
- \frac{D-2}{8(D-1)}\phi^2 R \sqrt{-g} 
= - \frac12 \partial_{\mu} \tilde{\phi} \partial_{\nu} \tilde{\phi} \tilde{g}^{\mu\nu} \sqrt{-\tilde{g}} 
- \frac{D-2}{8(D-1)}\tilde{\phi}^2 \tilde{R} \sqrt{-\tilde{g}} \;.
\label{L_matter}
\eea
Taking $\Omega = a$ we can simply work with the conformally rescaled metric 
\be
\tilde{g}_{\mu\nu} = \eta_{\mu\nu} + \kappa h_{\mu\nu}\;.
\ee
of which the inverse and the volume element are expanded as 
\bea
\tilde{g}^{\mu\nu} &=& 
\eta ^{\mu\nu} - \kappa h^{\mu\nu} + \kappa^2 h^{\mu}_{\phantom{\rho}\rho}h^{\rho\nu} 
+ \mathcal{O}(\kappa^3)\;, 
\\
\sqrt{-\tilde{g}} &=& 
1 + \frac{1}{2}\kappa h + \frac{1}{8}\kappa^2 h^2 - \frac{1}{4}\kappa^2 h^{\rho\sigma}h_{\rho\sigma}
+ \mathcal{O}(\kappa^3)\;.
\label{sqrt_g}
\eea
The next step is using the above expressions for the metric in order to expand the Ricci scalar at second order in $\kappa$. The expression for $\tilde{R}$ is,
\begin{eqnarray}
\tilde{R} = \kappa \Bigl(-h^{,\mu}_{\;\mu} + h^{\mu\nu}_{\phantom{\rho\rho},\mu\nu}\Bigr)
+\kappa^2 \Bigl(
-2 h^{\mu\nu} h_{\lambda\nu ,\mu}^{\phantom{\rho\rho\rho\rho}\lambda} 
+ h^{\mu\nu} h_{\mu\nu,\lambda}^{\phantom{\rho\rho\rho\rho}\lambda}  
+ h^{\mu\nu} h_{,\mu\nu} +\frac34 h^{\mu\nu}_{\phantom{\rho\rho},\lambda} h_{\mu\nu}^{\phantom{\rho\rho},\lambda} +h^{\mu\nu}_{\phantom{\rho\rho},\mu} h_{,\nu} 
-h^{\mu\nu}_{\phantom{\rho\rho},\mu} h_{\lambda\nu}^{\phantom{\rho\rho},\lambda}
\nonumber\\
\hspace{1.5cm}  
-\frac12 h^{\mu\nu}_{\phantom{\rho\rho},\lambda} h^{\lambda}_{\phantom{\rho}\nu ,\mu} 
-\frac14 h_{,\mu}h^{,\mu} 
-\frac12 h h^{,\mu}_{\;\mu} 
+\frac12 h h^{\mu\nu}_{\phantom{\rho\rho},\mu\nu}
\Bigr)\;+ \mathcal{O}(\kappa^3) \;.
\label{expandedRicci}
\end{eqnarray} 

\begin{figure}[htbp]
\centering
\includegraphics[height=2cm]{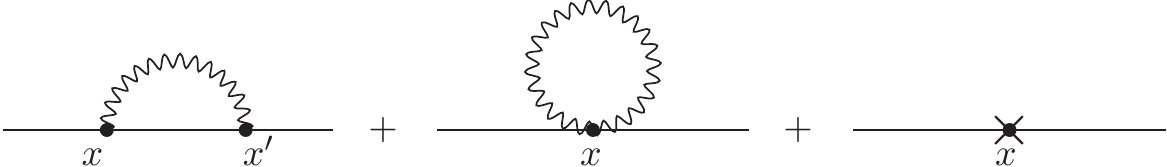}
\caption{The one-loop self-mass-squared from massless conformally coupled scalar. 
The first two diagrams represent the 3-point and 4-point interactions, respectively. 
The last diagram corresponds to the counterterms required 
to absorb ultraviolet divergences from the two primitive diagrams.}
\label{3diagrams}
\end{figure}

Using the perturbative expansion, the self-mass-squared $-iM^2(x;x')$ can be computed at any desired loop order. Our aim is to derive it at one loop order, which consists of the three Feynman diagrams depicted in Figure \ref{3diagrams}.
The first two diagrams which represent the 3-point and 4-point interactions, respectively 
have the following analytic expressions.
\bea
\mbox{1st diagram:} \quad -iM^2_{\rm 3pt}(x;x') &\equiv& 
\Bigl\langle i\frac{\delta S[\phi,h]}{\delta\phi(x)}i\frac{\delta S[\phi,h]}{\delta\phi(x')}\Bigr\rangle_0\;,
\label{M^2_3point}
\\
\mbox{2nd diagram:} \quad -iM^2_{\rm 4pt}(x;x') &\equiv& 
\Bigl\langle i\frac{\delta^2 S[\phi,h]}{\delta\phi(x)\delta\phi(x')} \Bigr\rangle_0\;,
\label{M^2_4point}
\eea
where the subscript $0$ on the expectation value indicates that it is to be taken in the free theory.
Once the action $S=\int d^Dx \mathcal{L}$ is expanded 
computing these two expressions is straightforward. 
The 3-point and 4-point diagrams correspond to 
the $\tilde{\phi}^2 h$ and $\tilde{\phi}^2 h^2$ interactions which derive from expanding 
the matter part of the Lagrangian \eqref{L_matter}. 
The expansion of the scalar kinetic term is given in the previous work \cite{kahya}
and therefore we focus on the conformal coupling term in \eqref{L_matter}. 
\begin{eqnarray}
\mathcal{L}_{\rm CC} &\equiv& -\frac{D-2}{8(D-1)}\tilde{\phi}^2 \tilde{R} \sqrt{-\tilde{g}}
\nonumber\\
&=& -\frac{D-2}{8(D-1)} \kappa\tilde{\phi}^2 \Bigl(-h^{,\mu}_{\;\mu} + h^{\mu\nu}_{\phantom{\rho\rho},\mu\nu}\Bigr)
-\frac{D-2}{8(D-1)}\kappa^2 \tilde{\phi}^2 \Bigl(
-2 h^{\mu\nu} h_{\lambda\nu ,\mu}^{\phantom{\rho\rho\rho\rho}\lambda} 
+ h^{\mu\nu} h_{\mu\nu,\lambda}^{\phantom{\rho\rho\rho\rho}\lambda}  
+ h^{\mu\nu} h_{,\mu\nu} 
\nonumber\\
& & \hspace{1.5cm}
+\frac34 h^{\mu\nu}_{\phantom{\rho\rho},\lambda} h_{\mu\nu}^{\phantom{\rho\rho},\lambda} 
+h^{\mu\nu}_{\phantom{\rho\rho},\mu} h_{,\nu} 
-h^{\mu\nu}_{\phantom{\rho\rho},\mu} h_{\lambda\nu}^{\phantom{\rho\rho},\lambda} 
-\frac12 h^{\mu\nu}_{\phantom{\rho\rho},\lambda} h^{\lambda}_{\phantom{\rho}\nu ,\mu} 
-\frac14 h_{,\mu}h^{,\mu} 
-\frac12 h h^{,\mu}_{\mu}
+\frac12 h h^{\mu\nu}_{\phantom{\rho\rho},\mu\nu}
\Bigr)\;.
\label{expandedLcc}
\end{eqnarray}

In the following subsection we derive the formal expressions for the self-mass-squared 
from the 3-point and 4-point interactions in the conformal coupling term. 
For notational simplicity we drop $\tilde{}~$ from now on, but remember that  
our metric, the scalar field $\phi$ and graviton field $h_{\mu\nu}$ are all conformally rescaled ones:
\bea
\tilde{g}_{\mu\nu} = \eta_{\mu\nu} + \kappa h_{\mu\nu}\;, \qquad
\tilde{\phi} 
\equiv \phi\;.
\label{rescaling_notation}
\eea

\subsection{Formal expressions for the one loop self-mass-squared}

\subsubsection{4-point contributions}

The 4-point contributions to the self-mass-squared come from $\phi^2 h^2$ terms 
in \eqref{expandedLcc}. 
Let us illustrate the derivation by calculating it from the first term,
\begin{equation}
\mathcal{L}_{4a} \equiv -\frac{D-2}{8(D-1)}\kappa^2 \phi^2 
(-2 h^{\mu\nu} h_{\lambda\nu ,\mu}^{\phantom{\rho\rho\rho\rho}\lambda})\;.
\end{equation} 
From the defining expression \eqref{M^2_4point}, the first step is to vary the action with respect to the scalar field,
\begin{eqnarray}
\frac{\delta^2 S_{4a}}{\delta\phi(x')\delta\phi(x)} 
&=& -\frac{D-2}{8(D-1)}\frac{\delta}{\delta\phi(x')} \int d^Dy 
    \Bigl[-2\kappa^2 \frac{\delta \phi^2(y)}{\delta \phi(x)} h^{\mu\nu}(y) 
    \partial^{\lambda} \partial_{\mu} h_{\mu\nu}(y))\Bigr]\;, 
\nonumber\\ 
&=&-\frac{D-2}{8(D-1)}
\Bigl[- 4\kappa^2 \delta^D(x-x') h^{\mu\nu}(x) \partial^{\lambda} \partial_{\mu} h_{\mu\nu}(x)\Bigr]\;,
\nonumber \\
&=& -\frac{D-2}{8(D-1)}\Bigl[- 4\kappa^2 \delta^D(x-x') h^{\mu\nu}(x')\partial^{\lambda} \partial_{\mu} h_{\mu\nu}(x)\Bigr]
\;. 
\label{4ptdif}
\end{eqnarray}
In the last line we used the delta function to convert $h^{\mu\nu}(x)$ to $h^{\mu\nu}(x')$ in order to distinguish them from one another. It also is useful to see the tensorial structure of the graviton propagator from $x$ to $x'$. The self-mass-squared is the functional integral of $i$ times (\ref{4ptdif}) over the relevant fields,
\begin{eqnarray}
-iM^2_{4a} (x;x') &\equiv& 
\Bigl\langle i\frac{\delta^2 S_{4a}}{\delta\phi(x')\delta\phi(x)} \Bigr\rangle_0\;,
\nonumber\\
&=& \frac{D-2}{2(D-1)} \kappa^2 i \delta^D(x-x') \partial^{\lambda} \partial_{\mu} 
 \Bigl\langle h^{\mu\nu}(x) h_{\lambda\nu}(x') \Bigr\rangle_0\;,
\nonumber\\
&=& \frac{32(D-1)}{(D-2)} \tilde{\kappa}^2 i \delta^D(x-x') \partial^{\lambda} \partial_{\mu} \Bigl\{
i[^{\mu\nu}\Delta_{\lambda\nu}](x;x')\Bigr\}\;.
\end{eqnarray}
Here we define $\tilde{\kappa} \equiv \frac{D-2}{8(D-1)}\times \kappa$ 
for future convenience.
Also note that an expression for the graviton propagator emerges in the last step:
\be
i[^{\mu\nu}\Delta_{\lambda\nu}](x;x') = 
\Bigl\langle h^{\mu\nu}(x) h_{\lambda\nu}(x') \Bigr\rangle_0\;.
\ee
Using the same procedure for the remaining terms in \eqref{expandedLcc} we have the contributions 
from the 4-point vertices as
\begin{eqnarray}
\lefteqn{-iM^2_{\rm 4pt}(x;x') = \frac{32(D-1)}{(D-2)} \tilde{\kappa}^2 i \delta^D(x-x')
\Biggl\{\;\partial^{\lambda} \partial'_{\mu} i[^{\mu\nu}\Delta_{\lambda\nu}](x;x')
- \frac12 \partial^{\lambda} \partial'_{\lambda} i[^{\mu\nu}\Delta_{\mu\nu}](x;x')
-\frac12 \partial_{\mu} \partial'_{\nu} i[^{\mu\nu}\Delta^{\alpha}_{\phantom{\rho}\alpha}](x;x')
} 
\nonumber\\
& & \hspace{2cm}
-\frac38 \partial^{\lambda} \partial'_{\lambda} i[^{\mu\nu}\Delta_{\mu\nu}](x;x') 
-\frac12 \partial_{\mu} \partial'_{\nu}i[^{\alpha}_{\phantom{\rho}\alpha}\Delta^{\mu\nu}](x;x')
+ \frac12\partial^{\lambda} \partial'_{\mu}i[^{\mu\nu}\Delta_{\lambda\nu}](x;x')
+\frac14 \partial^{\mu} \partial'_{\lambda} i[^{\lambda\nu}\Delta_{\mu\nu}](x;x')
\nonumber \\
& & \hspace{2cm}
+ \frac18 \partial_{\mu}\partial'^{\mu} i[^{\alpha}_{\phantom{\rho}\alpha}\Delta^{\beta}_{~\beta}](x;x')  
+\frac14 \partial^{\lambda} \partial'_{\lambda} i[^{\alpha}_{\phantom{\rho}\alpha}\Delta^{\beta}_{\phantom{\rho}\beta}](x;x')
- \frac14 \partial_{\mu}\partial'_{\nu}i[^{\alpha}_{\phantom{\rho}\alpha}\Delta^{\mu\nu}](x;x')\Biggl\}\;. \label{4-pt} 
\end{eqnarray}
\subsubsection{3-point contributions}

The 3-point contributions derive from $\phi^2 h$ terms in \eqref{expandedLcc}: 
\begin{eqnarray}
\mathcal{L}_{\rm 3pt} = 
-\frac{D-2}{8(D-1)} \kappa\phi^2 (-h^{,\mu}_{\;\mu} + h^{\mu\nu}_{\phantom{\rho\rho},\mu\nu})  = 
- \tilde{\kappa} \phi^2 (-h^{,\mu}_{\;\mu} + h^{\mu\nu}_{\phantom{\rho\rho},\mu\nu})
\equiv \mathcal{L}_{3a} + \mathcal{L}_{3b}\;.
\label{3pntlagr}
\end{eqnarray}
The 3-point self-mass-squared defined in \eqref{M^2_3point} then can be written as
\bea
-iM^2_{\rm 3pt}(x;x') = 
\Biggl\langle
i\frac{\delta S_{3a}}{\delta\phi(x)}i\frac{\delta S_{3a}}{\delta\phi(x')}
+i\frac{\delta S_{3a}}{\delta\phi(x)}i\frac{\delta S_{3b}}{\delta\phi(x')}
+i\frac{\delta S_{3b}}{\delta\phi(x)}i\frac{\delta S_{3a}}{\delta\phi(x')}
+i\frac{\delta S_{3b}}{\delta\phi(x)}i\frac{\delta S_{3b}}{\delta\phi(x')}
\Biggr\rangle_0\;.
\label{M^2_3point_expand}
\eea
Here the variations of the action with respect to the scalar are
\bea
\frac{\delta S_{3a}}{\delta\phi(x)} 
&=& - \tilde{\kappa} \frac{\delta}{\delta\phi(x)} \int d^Dy
\Bigl[-\phi^2(y)\partial_{\mu}\partial^{\mu} h(y)\Bigr] 
= - \tilde{\kappa} \Bigl[-2\phi(x)\partial_{\mu}\partial^{\mu} h(x)\Bigr]\;,
\label{S_3a}
\\
\frac{\delta S_{3b}}{\delta\phi(x)} 
&=& - \tilde{\kappa} \frac{\delta}{\delta\phi(x)} \int d^Dy
\Bigl[\phi^2(y)\partial_{\mu}\partial_{\nu}h^{\mu\nu}(y)\Bigr] 
= - \tilde{\kappa} \Bigl[2\phi(x)\partial_{\mu}\partial_{\nu}h^{\mu\nu}(x)\Bigr]\;.
\label{S_3b}
\eea
By substituting \eqref{S_3a}, \eqref{S_3b} into \eqref{M^2_3point_expand}, we obtain 
\begin{eqnarray} 
-iM^2_{\rm 3pt}(x;x')
= \!- \tilde{\kappa}^2 i\Delta_{\rm cf} \Biggl\{\!
\partial^2 \partial'^2 \; i[^{\alpha}_{\phantom{\rho}\alpha}\Delta^{\beta}_{\phantom{\rho}\beta}](x;x')
\!-\![\partial_{\mu} \partial_{\nu} \partial'^2 
\!+\!\partial'_{\mu} \partial'_{\nu} \partial^2 ] i[^{\alpha}_{\phantom{\rho}\alpha}\Delta^{\mu\nu}](x;x')
\!+\!\partial_{\mu} \partial_{\nu} \partial'_{\alpha} \partial'_{\beta} i[^{\mu\nu}\Delta^{\alpha\beta}](x;x')\!\Biggr\}\;.
\label{3-pt}
\end{eqnarray}  
 Here note that the appearance of the propagator for a massless, conformally coupled scalar 
\be
i\Delta_{\rm cf}(x;x') = 
\Bigl\langle \phi(x)\phi(x') \Bigr\rangle_0\;.
\ee

In fact, there exists another kind of 3-point contributions 
taking one variation from the kinetic term and the other from the conformal coupling 
in the Lagrangian \eqref{L_matter}. The 3-point interactions from the scalar kinetic term
are
\be
\mathcal{L}_{\rm K 3pt} = -\frac{1}{2}\kappa\partial_{\mu}\phi\partial_{\nu}\phi
\Bigl(- h^{\mu\nu} +\frac{1}{2}\eta^{\mu\nu} h\Bigr) 
\equiv \mathcal{L}_{K3c} + \mathcal{L}_{K3d}\;,
\ee 
and the 3-point self-mass-squared from the `cross terms' can be computed as 
\bea
-iM^2_{\rm 3pt cross}(x;x') = 
\Biggl\langle
i\frac{\delta S_{3a}}{\delta\phi(x)}i\frac{\delta S_{K3c}}{\delta\phi(x')}
+i\frac{\delta S_{3a}}{\delta\phi(x)}i\frac{\delta S_{K3d}}{\delta\phi(x')}
+i\frac{\delta S_{3b}}{\delta\phi(x)}i\frac{\delta S_{K3c}}{\delta\phi(x')}
+i\frac{\delta S_{3b}}{\delta\phi(x)}i\frac{\delta S_{K3d}}{\delta\phi(x')}
\Biggr\rangle_0\;.
\label{M^2_3point_cross}
\eea   
Including \eqref{3-pt}, \eqref{M^2_3point_cross} and the contributions from the kinetic terms given 
in \cite{kahya} (which can be formally written as)
\bea
-iM^2_{\rm 3pt K}(x;x') = 
\Biggl\langle
i\frac{\delta S_{K3c}}{\delta\phi(x)}i\frac{\delta S_{K3c}}{\delta\phi(x')}
+i\frac{\delta S_{K3c}}{\delta\phi(x)}i\frac{\delta S_{K3d}}{\delta\phi(x')}
+i\frac{\delta S_{K3d}}{\delta\phi(x)}i\frac{\delta S_{K3c}}{\delta\phi(x')}
+i\frac{\delta S_{K3d}}{\delta\phi(x)}i\frac{\delta S_{K3d}}{\delta\phi(x')}
\Biggr\rangle_0\;.
\label{M^2_3point_K}
\eea   
completes the 3-point contributions. However, with the experience from \cite{kahya} and the present work, each of the three expressions \eqref{3-pt}, \eqref{M^2_3point_cross} and \eqref{M^2_3point_K} takes a number of heavy computations and can be treated separately. Thus we leave calculating the cross contributions as a follow-up project.

\subsection{Propagators}

From the 4-point and 3-point contributions \eqref{4-pt} and \eqref{3-pt}, 
we see that the propagators are the main ingredients of the the self-mass-squared.
This subsection therefore provides with the scalar and graviton propagators.  
We follow three notational conventions employed in the previous work \cite{kahya} 
for continuity and comparison with the current results. 
First, the background geometry is denoted with a hat, 
which was already introduced in \eqref{bkgd_geometry},
\be
\hat{g}_{\mu\nu} = a^2\eta_{\mu\nu} \quad \mbox{accordingly} \quad \hat{R} = D (D-1) H^2\;.
\ee
Second, considering the gauge choice for the graviton propagator 
in which time and space are treated differently, it is convenient to differentiate 
the spatial parts of the Lorentz metric and the Kronecker delta with a bar, 
\begin{equation}
\overline{\eta}_{\mu\nu} \equiv \eta_{\mu\nu} + \delta^0_{\mu} \delta^0_{\nu} \qquad {\rm and} \qquad \overline{\delta}^{\mu}_{\nu} \equiv \delta^{\mu}_{\nu} - \delta_0^{\mu} \delta^0_{\nu}\;.
\end{equation}
Third, it is useful to define the de Sitter length function $y(x;x')$
in terms of the de Sitter invariant length $\ell(x;x')$ from $x^{\mu}$ to $x^{\prime \mu}$:
\begin{eqnarray}\label{ydef}
y(x;x')
=4 \sin^2\Bigl( \frac12 H \ell(x;x')\Bigr) 
= a a' H^2 \Bigl\{ \Vert \vec{x} - \vec{x}' \Vert^2 - \Bigl(\vert \eta - \eta'\vert - i \delta\Bigr)^2 \Bigr\}\;,
\end{eqnarray}
where $a \equiv a(\eta)$ and $a' \equiv a(\eta')$.
The propagator for a massless conformally coupled scalar has long been known \cite{BD},
\begin{equation}
{i\Delta}_{\rm cf}(x;x') = \frac{H^{D-2}}{(4\pi)^{\frac{D}2}}
\Gamma\Bigl( \frac{D}2 \!-\! 1\Bigr)
\Bigl(\frac4{y}\Bigr)^{\frac{D}2-1} \; . \label{CCP}
\end{equation}

The graviton propagator was derived by adding the gauge fixing term to the Lagrangian 
\cite{tw},
\bea
\label{gf}
\mathcal{L}_{\rm GF} = -\frac12 a^{D-2} \eta^{\mu\nu} F_{\mu} F_{\nu}\;, 
\quad
F_{\mu} & \equiv & \eta^{\rho\sigma} \Bigl(h_{\mu\rho ,\sigma} 
- \frac12 h_{\rho \sigma ,\mu} + (D-2) H a h_{\mu \rho} \delta^0_{\sigma} \Bigr)\;.
\eea
The quadratic part of the gauge fixed Lagrangian can be partially integrated to extract
the kinetic operator $D_{\mu\nu}^{~~\rho \sigma}$ as
\be
\frac12 h^{\mu\nu} D_{\mu\nu}^{~~\rho \sigma} h_{\rho\sigma}\;,
\ee
where
\begin{eqnarray}
D_{\mu\nu}^{\phantom{\rho\rho}\rho\sigma}&\equiv &\Biggl\{ \frac12 \overline{\delta}_{\mu}^{\phantom{\rho}(\rho} \overline{\delta}_{\nu}^{\phantom{\rho}\sigma)} - \frac14 \eta_{\mu\nu} \eta^{\rho\sigma}-\frac1{2(D-3)} \delta_{\mu}^0 \delta_{\nu}^0 \delta_0^{\rho} \delta_0^{\sigma} \Biggr\} D_A 
+\delta^0_{(\mu} \overline{\delta}_{\nu )}^{( \rho} \delta_0^{\sigma )} D_B + \frac12 \Bigl(\frac{D-2}{D-3}\Bigr) \delta_{\mu}^0 \delta_{\nu}^0 \delta_0^{\rho} \delta_0^{\sigma} D_C\;.
\end{eqnarray}
The three scalar differential operators are defined as
\begin{eqnarray}
D_A & \equiv & \partial_{\mu} \Bigl(\sqrt{-\hat{g}} \,
\hat{g}^{\mu\nu} \partial_{\nu}\Bigr) \;, \\
D_B & \equiv & \partial_{\mu} \Bigl(\sqrt{-\hat{g}} \,
\hat{g}^{\mu\nu} \partial_{\nu}\Bigr) - \frac1{D} 
\Bigl(\frac{D\!-\!2}{D\!-\!1}\Bigr) \hat{R} 
\sqrt{-\hat{g}} \; , \\
D_C & \equiv & \partial_{\mu} \Bigl(\sqrt{-\hat{g}} \,
\hat{g}^{\mu\nu} \partial_{\nu}\Bigr) - \frac2{D}
\Bigl(\frac{D\!-\!3}{D\!-\!1}\Bigr) \hat{R} 
\sqrt{-\hat{g}} \; .
\end{eqnarray}
The graviton propagator should obey the following defining equation,
\begin{equation}
D_{\mu\nu}^{~~\rho\sigma} \times i\Bigl[{}_{\rho\sigma}
\Delta^{\alpha\beta} \Bigr](x;x') = \delta_{\mu}^{(\alpha}
\delta_{\nu}^{\beta)} i \delta^D(x-x') \;,
\end{equation}  
and by inverting the kinetic operator one can get the graviton propagator. 
This inversion procedure can be done for each scalar kinetic operator
by setting the graviton propagator as a sum of constant tensor factors times scalar propagators:
\begin{equation}
i[_{\mu\nu}\Delta_{\rho\sigma}](x;x') = \underset{I =A, B, C}{\Sigma} [_{\mu\nu} T^I_{\rho\sigma}] i \Delta_I(x;x')\;,
\end{equation}
where the scalar propagators satisfy
\begin{equation}
D_I \times i\Delta_I(x;x') = i \delta^D(x - x') \qquad {\rm for}
\qquad I = A,B,C \;. \label{sprops}
\end{equation}
Here the tensor factors are given as
\begin{eqnarray}
\Bigl[{}_{\mu\nu} T^A_{\rho\sigma}\Bigr] &=&  2 \overline{\eta}_{\mu (\rho} \overline{\eta}_{\sigma) \nu} -
\frac{2}{D-3} \overline{\eta}_{\mu\nu} \overline{\eta}_{\rho \sigma},\label{T^A} \\
\Bigl[{}_{\mu\nu} T^B_{\rho\sigma}\Bigr] &=& -4 \delta^0_{(\mu} \overline{\eta}_{\nu) (\rho} \delta^0_{\sigma)}, \label{T^B} \\
\Bigl[{}_{\mu\nu} T^C_{\rho\sigma}\Bigr] &=&  \frac{2}{(D-3)(D-2)} 
[(D-3) \delta^0_{\mu} \delta^0_{\nu} + \overline{\eta}_{\mu\nu}] 
[(D-3)\delta^{0}_{\rho}\delta^{0}_{\sigma} + \overline{\eta}_{\rho\sigma}]. \label{T^C}
\end{eqnarray}

The $A$-type propagator which is the same as the one for the MMC scalar 
has de Sitter invariant (as a function of $y$) and breaking parts \cite{ow1,ow2}.
\begin{eqnarray}
i \Delta_A (x;x') &=& A(y) + k \ln(aa')\;, 
\end{eqnarray}
where $k \equiv \frac{H^{D-2}}{(4 \pi )^{D/2}} \frac{\Gamma(D-1)}{\Gamma(\frac{D}2)}$.
The de Sitter invariant part $A(y)$ is \cite{ow2},
\begin{eqnarray}\label{A}
\lefteqn{A(y)\equiv \frac{H^{D-2}}{(4\pi)^{D/2}} \Biggl\{ 
\Gamma(\frac{D}2 -1)\Bigl(\frac{4}{y}\Bigr)^{\frac{D}2 -1} 
+\frac{\Gamma(\frac{D}2 +1)}{\frac{D}2 -2} \Bigl(\frac{4}{y} \Bigr)^{\frac{D}2 -2}
-\pi \cot\Bigl(\frac{\pi D}2\Bigr) \frac{\Gamma(D-1)}{\Gamma(\frac{D}2)}
}
\nonumber \\
&& \hspace{2.5cm} +\sum_{n=1}^{\infty}\biggl[ 
\frac1{n} \frac{\Gamma(n+D-1)}{\Gamma(n+\frac{D}2)} \Bigl(\frac{y}4 \Bigr)^n 
-\frac1{n -\frac{D}2 +2} \frac{\Gamma(n +\frac{D}2 +1)}{ \Gamma(n+2)}\Bigl(\frac{y}4 \Bigr)^{n -\frac{D}2 +2} \biggr] \Biggr\}\;. \label{DeltaAA} 
\end{eqnarray}
Note that this de Sitter breaking solution still preserves homogeneity and isotropy 
and it is a well-known issue that there is no de Sitter invariant solution for 
the MMC scalar propagator \cite{af}.

The $B$-type and $C$-type propagators have the following de Sitter invariant solutions 
\begin{eqnarray}
i \Delta_B (x;x') &\!\!\!=\!\!\!& 
i \Delta_{\rm cf}(x;x') 
-\frac{H^{D-2}}{(4\pi )^{D/2}} \sum_{n=0}^{\infty} \Biggl\{ 
\frac{\Gamma(n+D\!-\!2)}{\Gamma (n+\frac{D}2)}\Bigl(\frac{y}4 \Bigr)^n 
-\frac{\Gamma (n\!+\!\frac{D}2)}{\Gamma (n\!+\!2)} \Bigl(\frac{y}4 \Bigr)^{n \!-\!\frac{D}2 \!+\!2}
\Biggr\} \;, 
\label{DeltaB} 
\\
i \Delta_C (x;x') &\!\!\!=\!\!\!& 
i \Delta_{\rm cf}(x;x')
+\frac{H^{D-2}}{(4\pi )^{D/2}} \sum_{n=0}^{\infty} \Biggl\{ 
(n\!+\!1)\frac{\Gamma(n\!+\!D\!-\!3)}{\Gamma (n\!+\!\frac{D}2)}\Bigl(\frac{y}4 \Bigr)^n 
-\Bigl(n\!-\!\frac{D}2\!+\!3\Bigr)\frac{\Gamma(n\!+\!\frac{D}2 \!-\!1)}{\Gamma(n+2)}
\Bigl(\frac{y}4 \Bigr)^{n\!-\!\frac{D}2\!+\!2}
 \Biggr\}\;. 
\label{DeltaC} 
\end{eqnarray}
The infinite series terms of a positive power of $y$ vanish for $D=4$ so that one only need to retain them when multiplying a fixed divergence. This makes these propagators and loop calculations manageable.   

\section{Computation of the one loop self-mass-squared}\label{ccssm}

In this section we evaluate the formal expressions for 
the 4-point and 3-point contributions to the self-mass-squared
given in \eqref{4-pt} and \eqref{3-pt}. 
The contribution from the 4-point vertices 
turns out to be finite.
To manage the divergences occurring in the 3-point interactions 
we put them in the form of external operators acting on functions of $y$. 
This form is convenient for renormalization in the next section.     

\subsection{Contribution from the 4-point vertices}

\begin{figure}[htbp]
\centering
\includegraphics[height=3cm]{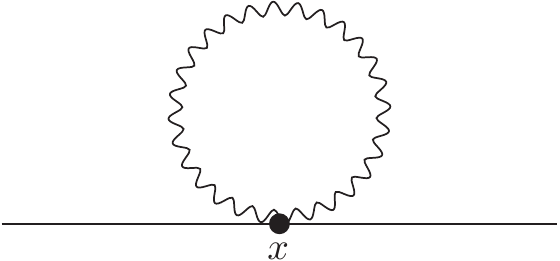}
\caption{Contribution from the 4-point vertices.}
\label{4pt}
\end{figure}

We start with the analytic expression for the 4-point contribution $-iM^2_{\rm 4pt}(x;x')$ in \eqref{4-pt} 
corresponding to the following Feynman diagram, Fig. \ref{4pt}.

We take the coincidence limit with the aid of the delta function. 
Again we work it out for the first term to demonstrate the procedure,
\begin{equation}
\delta^D(x-x')\partial^{\lambda}\partial_{\mu} i[^{\mu\nu}\Delta_{\lambda\nu}](x;x') 
= \lim_{x\rightarrow x'}\partial^{\lambda}\partial_{\mu} i[^{\mu\nu}\Delta_{\lambda\nu}](x;x')\;.
\end{equation}
The constant tensor factors of this term are evaluated from \eqref{T^A}, \eqref{T^B} and \eqref{T^C},
\begin{eqnarray}
\Bigl[{}^{\mu\nu} T^A_{\lambda\nu} \Bigr] &=& \frac{D^2 -3D -2}{D-3} \overline{\delta}^{\mu}_{\lambda}\;, \\ 
\Bigl[{}^{\mu\nu} T^B_{\lambda\nu} \Bigr] &=& \overline{\delta}^{\mu}_{\lambda} + (D-1) \delta^{\mu}_{0}\delta^{0}_{\lambda}\;, \\
\Bigl[{}^{\mu\nu} T^C_{\lambda\nu} \Bigr] &=& \frac{2}{(D-3)(D-2)}[(D-3)^2 \delta^{\mu}_{0} \delta^{0}_{\lambda} + \overline{\delta}^{\mu}_{\lambda}]\;. 
\end{eqnarray}
and the differentiated scalar propagators become
\begin{eqnarray}
\lim_{x \rightarrow x'} \partial^{\lambda}\partial_{\mu}i[\Delta_A](x;x')
&=& \frac{H^{D}a^2}{(4\pi)^{D/2}} \frac{\Gamma (D-1)}{\Gamma (\frac{D}{2})} \Biggl\{\frac{D-1}{D} \delta^{\lambda}_{\mu} - \delta^{\lambda}_{0} \delta^{0}_{\mu} \Biggr\}\;, 
\\
\lim_{x \rightarrow x'} \partial^{\lambda}\partial_{\mu}i[\Delta_B](x;x')
&=& \frac{H^{D}a^2}{(4\pi)^{D/2}} \frac{\Gamma( D-1)}{\Gamma (\frac{D}{2})} \Biggl\{-\frac{1}{D} \delta^{\lambda}_{\mu} \Biggr\}\;, 
\\
\lim_{x \rightarrow x'} \partial^{\lambda}\partial_{\mu}i[\Delta_C](x;x')
&=& \frac{H^{D}a^2}{(4\pi)^{D/2}} \frac{\Gamma (D-1)}{\Gamma (\frac{D}{2})} \Biggl\{\frac{2}{D(D-2)} \delta^{\lambda}_{\mu} \Biggr\}\;.
\end{eqnarray}
Therefore for the first term we have,
\begin{eqnarray}
\lim_{x\rightarrow x'} \partial^{\lambda} \partial_{\mu} [^{\mu\nu} \Delta_{\lambda\nu}](x;x')
&=& \frac{H^{D}a^2}{(4\pi)^{D/2}}\frac{\Gamma (D-1)}{\Gamma (\frac{D}{2})} \frac{(D^4-6D^3+9D^2+4D-16)}{(D-2)^2}\;, \nonumber\\
&\rightarrow& \frac{H^{4}a^2}{16\pi^2} \times 8 \qquad \mbox{in }~ D=4\;. 
\label{first}
\end{eqnarray}
Using the same procedure for the remaining nine terms we obtain the 4-point 
contributions in $D=4$ dimensions,
\begin{eqnarray}
-iM^2_{\rm 4pt}(x;x') &=&  \frac{i \tilde{\kappa}^2}{3} \delta^4(x-x') \frac{H^{4}a^2}{16\pi^2} 
\times \Biggl\{1 \times (8) -\frac12 \times (20) -\frac12 \times (-16) -\frac38 \times (-32)
\nonumber\\
& & \hspace{1cm}
-\frac12 \times (16)  +\frac12 \times (-8) +\frac14 \times (-8) +\frac18 \times (64)
+\frac14 \times (-40) -\frac14 \times (-16) \Biggr\}\;, 
\nonumber\\
&=& 2 i \kappa^2 \times \frac{H^{4}a^2}{16\pi^2} \delta^4 (x-x')\;. 
\label{4pt-4dim}
\end{eqnarray}

\subsection{3-point interaction}
\begin{figure}[htbp]
\centering
\includegraphics[height=2.4cm]{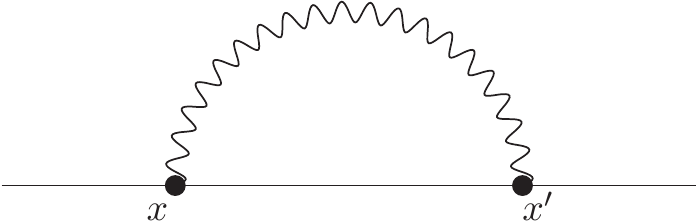}
\caption{Contribution from the 3-point vertices.}
\label{3pt}
\end{figure}

In this subsection we evaluate the 3-point contribution $-iM^2_{\rm 3pt}(x;x')$ in \eqref{3-pt} 
depicted in the following diagram, Fig. \ref{3pt}.

Substituting the graviton propagator and contracting the tensor indices in \eqref{3-pt} we have
\begin{eqnarray} 
= \lefteqn{-\tilde{\kappa}^2 i\Delta_{\rm cf} \Biggl\{
- \frac{4(D-1)}{D-3}\partial^2\partial'^2 i\Delta_{A} 
+ \frac{8}{(D-3)(D-2)} \partial^2 \partial'^2 i\Delta_{C} 
-\frac4{D-3} \biggl[-\nabla^2 \partial'^2 i\Delta_{A}
}
\nonumber \\
& & \hspace{0.5cm}
-\nabla^2 \partial^2 i\Delta_{A}
+\frac1{D-2} \Bigl[(D-3)\partial_{0}^{2} + \nabla^2 \Bigr] \partial'^2 i\Delta_{C} 
+\frac1{D-2} \Bigl[(D-3)\partial_{0}^{'2}+\nabla^2 \Bigr] \partial^2 i\Delta_{C} \biggr]
+\frac{2(D-4)}{D-3}\nabla^4 i\Delta_{A} 
\nonumber \\
& & \hspace{0.5cm}
+ 4\nabla^2 \partial_0 \partial'_{0} i\Delta_{B} + \frac2{(D-3)(D-2)} \Bigl[(D-3)^2\partial_{0}^{2} \partial_{0}^{'2} +(D-3)\nabla^2 (\partial_{0}^{'2} + \partial_{0}^{2}) + \nabla^4 \Bigr ] i\Delta_{C} \Biggr\}\;.\label{Msqr3}
\end{eqnarray}
Here we applied 
\be
\partial'_i = -\partial_i\;, \quad
\nabla'^2 = \nabla^2\;, \quad 
\partial^2 = -\partial_{0}^{2} + \nabla^2\;.
\ee
Also, in the de Sitter background the covariant d'Alembertian operator is expressed as
\be
\square \equiv \frac1{\sqrt{-\hat{g}}} \partial_{\mu} \Bigl(\sqrt{-\hat{g}} \hat{g}^{\mu\nu} \partial_{\nu}\Bigr) = -\frac1{a^2} \partial_{0}^2 - \frac{(D-2)H}{a} \partial_{0}+\frac1{a^2}\nabla^2.
\ee
Let us remember that the expression for the conformal propagator was defined for de Sitter geometry in \eqref{CCP}. Since we conformally rescaled our metric to flat space and the CC scalar is invariant under this rescaling we can work with the flat space limit of $i\Delta_{\rm cf}$. In order to do the calculations more systematically we express all the propagators in terms of the de Sitter invariant function $y(x;x')$ in (\ref{ydef}),
\bea
i\Delta_{\rm cf} & = & \frac{\Gamma\Bigl( \frac{D}2 \!-\! 1\Bigr)}{(4\pi)^{\frac{D}2}}
\Bigl(\frac4{\Delta x^2}\Bigr)^{\frac{D}2-1} = (aa')^{\frac{D}{2}-1} \frac{H^{D-2}}{(4\pi)^{\frac{D}2}}
\Gamma\Bigl( \frac{D}2 \!-\! 1\Bigr)
\Bigl(\frac4{y}\Bigr)^{\frac{D}2-1} \equiv 
(aa')^{\frac{D}{2}-1} F(y)\;,   \\ 
i\Delta_{A, B, C} & \equiv & A(y), B(y), C(y)\;.
\eea
Obviously a similar flat space version for the graviton can not be used for this calculation since the graviton is not conformally invariant. Note also that the de Sitter breaking term in $i\Delta_{A}$ is dropped. Then the terms in the 3-point contribution can be classified as the following four types,
\begin{eqnarray}
F(y)\partial_{0}^{2} \partial_{0}^{'2}A(y),\;\; F(y)\nabla^4 A(y), \;\;
F(y)\nabla^2 [\partial_{0}^{2} + \partial_{0}^{'2}]\; A(y)\;\; {\rm and}\;\; F(y)\nabla^2 \partial_{0}\partial'_{0} A(y)\;.
\label{4-types}
\end{eqnarray}
This allows to write the 3-point contribution as
\begin{eqnarray}
\lefteqn{-iM^2_{\rm 3pt}(x;x') = 
-\tilde{\kappa}^2 (aa')^{\frac{D}{2}-1} \Biggl\{C_{1a} F(y)\partial_{0}^{2} \partial_{0}^{'2} A(y) 
+ C_{2a} F(y)\nabla^4 A(y) 
+ C_{3a} F(y)\nabla^2 [\partial_{0}^{2} + \partial_{0}^{'2}] A(y)
} 
\nonumber \\
& & \hspace{2cm} 
+ C_{b} F(y) \nabla^2 \partial_{0} \partial'_{0} B(y)
+ C_{1c} F(y) \partial_{0}^{2} \partial_{0}^{'2} C(y) + C_{2c} F(y)\nabla^4 C(y) 
+ C_{3c} F(y) \nabla^2 [\partial_{0}^{2} + \partial_{0}^{'2}] C(y)\!\Biggr\}\;. 
\label{3-pt-ABC}
\end{eqnarray}
Here the coefficients are 
\begin{eqnarray} \label{coeffsABC}
C_{1a} &\!\!\!\equiv\!\!\!& -4(\frac{D-1}{D-3}) \;, \; 
C_{2a} \equiv  -2(\frac{D-2}{D-3}) \;, \; 
C_{3a} \equiv  4(\frac{D-2}{D-3}) \;, \; 
C_{b\phantom{\eta}} \equiv 4 \;,
\nonumber\\
C_{1c} &\!\!\!\equiv\!\!\!& \frac{2(D-2)^2}{(D-3)(D-2)} \;, \;
C_{2c} \equiv  \frac2{(D-3)(D-2)}\; \;
{\rm and}\;\;C_{3c} \equiv \frac{(D-1)}{(D-3)(D-2)}\;. 
\end{eqnarray}

Recalling the self-mass-squared is eventually integrated over 4-dimensional spacetime
in the effective field equation \eqref{linear_eq}, we extract the derivatives outside the integral 
to reduce the degree of divergence of the terms remaining inside the integral.
This requires to convert all primed derivatives into unprimed ones 
(so that they can be freely moved outside the integral.)
The final results are expressed in terms of eleven external operators acting on the functions of 
$y$. 
The procedure of extracting derivatives involves a number of indefinite integration. 
We denote this operation by 
\begin{equation}
I[f](y) \equiv \int^y dy' f(y') \; .
\end{equation}
For example, the following identity derived in \cite{kahya} 
\begin{equation}
f(y) \Biggl\{A''(y) \frac{\partial y}{\partial x^{\rho}} \frac{\partial y}{
\partial x^{\prime \sigma}} \!+\! A'(y) \frac{\partial^2 y}{\partial x^{\rho}
\partial x^{\prime \sigma}} \Biggr\} \!=\!
\partial_{\rho} \partial_{\sigma}' I^2[f A''](y) \!+\! \frac{\partial^2 y}{
\partial x^{\rho} \partial x^{\prime \sigma}} I[f' A'](y) \; . \label{keyI}
\end{equation}
describes the operation: 
the expression in the left-hand-side is converted to derivatives acting a function of $y$ 
(as an indefinite integral function of $y$) plus an extra function of $y$ in the right-hand-side. 
Similar identities can be found in \cite{kahya}. 
\begin{eqnarray}
f(y) \partial_0 \partial_0' A(y) & = & \partial_0 \partial_0'
I^2[f A''](y) -\frac12 \nabla \!\cdot\! \nabla' I^3[f' A'](y) 
+ H^2 a a' \Bigl\{(2 \!-\! y) I[f'A'](y) -
(D\!-\!1) I^2[f' A'](y)\Bigr\} \; ,\label{ID013} \qquad \\
f(y) \partial_0 \partial_j' A(y) & = & \partial_0 \partial_j'
I^2[f A''](y) + H a \partial_j' I^2[f' A'](y) \; , \qquad \\
f(y) \partial_i \partial_0' A(y) & = & \partial_i \partial_0'
I^2[f A''](y) + H a' \partial_i I^2[f' A'](y) \; , \qquad \\
f(y) \partial_i \partial_j' A(y) & = & \partial_i \partial_j'
I^2[f A''](y) - 2 H^2 a a' \eta_{ij} I[f' A'](y) \; .
\label{ID016}\qquad
\end{eqnarray}
Note that the derivatives are acting on functions of $y$ externally in the above expressions. For our computation we need to derive these kind of expressions for a more complicated situation where there are four derivatives, spatial and temporal, acting on functions $A(y), B(y), C(y)$ that appear in equation \eqref{3-pt-ABC},
\begin{eqnarray}
\lefteqn{F(y)\partial_{\rho}\partial_{\sigma} \partial'_{\alpha}\partial'_{\beta} A(y) 
= \partial_{\rho} \partial_{\sigma} \partial'_{\alpha} \partial'_{\beta} \bigg\{I^4[FA''''](y) \bigg\} 
+ \frac{\partial^4 y}{\partial x^{\rho} \partial x^{\sigma} \partial x^{'\alpha} \partial x^{'\beta}}
\bigg\{I[F'A'] + I^2[F'A''] + I^3[F'A''']\bigg\}
} \nonumber\\
& & + \bigg(\!
2 \frac{\partial y}{\partial x^{(\rho}} \frac{\partial^3 y}{\partial x^{\sigma)}\partial x'^{\alpha}\partial x'^{\beta}} 
\!+\! 2 \frac{\partial y}{\partial x'^{(\alpha}} \frac{\partial^3 y}{\partial x'^{\beta)}\partial x^{\rho}\partial x^{\sigma}} 
\!+\! \frac{\partial^2 y}{\partial x^{\rho}\partial x^{\sigma}} \frac{\partial^2 y}{\partial x'^{\alpha}\partial x'^{\beta}} 
\!+\! 2\frac{\partial^2 y}{\partial x^{\rho}\partial x^{'(\alpha}} \frac{\partial^2 y}{\partial x'^{\beta)}\partial x^{\sigma}}
\!\bigg)\!\!\bigg\{\!I[F'A''] \!+\! I^2[F'A'''] \!\bigg\}  
\nonumber \\
& & + \bigg(\!4 
\frac{\partial y}{\partial x^{(\rho}} \frac{\partial^2 y}{\partial x^{\sigma)}\partial x'^{(\alpha}} \frac{\partial y}{\partial x'^{\beta)} } 
\!+\! \frac{\partial y}{\partial x^{\rho}}\frac{\partial y}{\partial x^{\sigma}} \frac{\partial^2 y}{\partial x'^{\alpha}\partial x'^{\beta}} 
\!+\! \frac{\partial y}{\partial x'^{\alpha}} \frac{\partial y}{\partial x'^{\beta}} \frac{\partial^2 y}{\partial x^{\rho} \partial x^{\sigma} }\!\bigg)\!\! \bigg\{I[F'A''']\bigg\}\;.
\end{eqnarray}  
Let us apply this method to the simplest cases, which is 
all of the derivatives are spatial, i.e. $F(y)\nabla^4 A(y)$ or $F(y)\nabla^4 C(y)$ terms in equation \eqref{3-pt-ABC}.
\begin{eqnarray}
\lefteqn{F(y)\nabla^4 A(y) 
= \partial_{i} \partial_{i} \partial'_{j} \partial'_{j} \bigg\{I^4[FA''''](y) \bigg\} 
+ \frac{\partial^4 y}{\partial x^{i} \partial x^{i} \partial x^{'j} \partial x^{'j}}
\bigg\{I[F'A'] + I^2[F'A''] + I^3[F'A''']\bigg\}
} \nonumber\\
& & + \bigg(\!
2 \frac{\partial y}{\partial x^{(i}} \frac{\partial^3 y}{\partial x^{i)}\partial x'^{j}\partial x'^{j}} 
\!+\! 2 \frac{\partial y}{\partial x'^{(j}} \frac{\partial^3 y}{\partial x'^{j)}\partial x^{i}\partial x^{i}} 
\!+\! \frac{\partial^2 y}{\partial x^{i}\partial x^{i}} \frac{\partial^2 y}{\partial x'^{j}\partial x'^{j}} 
\!+\! 2\frac{\partial^2 y}{\partial x^{i}\partial x^{'(j}} \frac{\partial^2 y}{\partial x'^{j)}\partial x^{i}}
\!\bigg)\!\!\bigg\{\!I[F'A''] \!+\! I^2[F'A'''] \!\bigg\}  
\nonumber \\
& & + \bigg(\!4 
\frac{\partial y}{\partial x^{(i}} \frac{\partial^2 y}{\partial x^{i)}\partial x'^{(j}} \frac{\partial y}{\partial x'^{j)} } 
\!+\! \frac{\partial y}{\partial x^{i}}\frac{\partial y}{\partial x^{i}} \frac{\partial^2 y}{\partial x'^{j}\partial x'^{j}} 
\!+\! \frac{\partial y}{\partial x'^{j}} \frac{\partial y}{\partial x'^{j}} \frac{\partial^2 y}{\partial x^{i} \partial x^{i} }\!\bigg)\!\! \bigg\{I[F'A''']\bigg\}\;. \label{nabla1}
\end{eqnarray}   
As one can see from the above equation we need to take up to four derivatives of $y$ with respect to spatial variable $x^i$ or $x'^{j}$ and make use of 
equation \eqref{ID016}. Using the following derivative identities
\begin{eqnarray}
\frac{\partial y}{\partial x^i} = 2 H^2 a a' \Delta x_i \; , \;\;
\frac{\partial y}{\partial x'^{j}} = - 2 H^2 a a' \Delta x_j \; , \;\;
\frac{\partial^2 y}{\partial x^i \partial x^{j}} = 2 H^2 a a' \eta_{i j} \; , \;\;
\frac{\partial^2 y}{\partial x^i \partial x'^{j}} = - 2 H^2 a a' \eta_{i j} \;, 
\nonumber \\ 
\frac{\partial^2 y}{\partial x^i \partial x^{i}} = \frac{\partial^2 y}{\partial x'^j \partial x'^{j}} = 2 H^2 a a' (D-1) \; , \;\;
\frac{\partial^3 y}{\partial x^i \partial x^i \partial x'^j} = \frac{\partial^3 y}{\partial x^i \partial x'^j \partial x'^j} = \frac{\partial^4 y}{\partial x^{i} \partial x^{i} \partial x^{'j} \partial x^{'j}} = 0\;.
\end{eqnarray} 
Eq. \eqref{nabla1} will reduce into
\bea
\lefteqn{F(y)\nabla^4 A(y) 
= \nabla^4 \bigg\{I^4[FA''''](y) \bigg\} 
+ \bigg( \Bigl[2 H^2 a a' (D-1)\Bigr]^2 + 2 \Bigl[2 H^2 a a'\Bigr]^2 (D-1) \bigg)\!\!  \bigg\{\!I[F'A''] \!+\! I^2[F'A'''] \!\bigg\}}  
\nn \\
&& + \bigg(\!4 
\Bigl[2 H^2 a a'\Bigr]^3 \Delta x_i \Delta x_j \eta_{i j} + \Bigl[2 H^2 a a'\Bigr]^3 \Delta x_i \Delta x_i (D-1) + \Bigl[2 H^2 a a'\Bigr]^3 \Delta x_i \Delta x_i (D-1) \bigg) \bigg\{I[F'A''']\bigg\}\;
\nn \\
&& \hspace{-0.3cm} = \nabla^4 \bigg\{I^4[FA''''](y) \bigg\}  + 8(D^2-1) \Bigl[H^2 a a'\Bigr]^2 \bigg\{I[F'A''] + I^2[F'A'''] \bigg\} + 16(D+1) \Bigl[H^2 a a'\Bigr]^3 \Vert \Delta \vec{x} \Vert^2 \bigg\{I[F'A''']\bigg\}\;.
\label{nabla2}
\eea
The final form of \eqref{nabla1} is:
\begin{eqnarray}
F(y)\nabla^4 A(y) 
= \nabla^4 \bigg\{I^4[FA''''](y) \bigg\} 
+ 4(D^2-1) H^4 (a a')^2 \bigg\{I[F'A''] - I^2[F'A'''] \bigg\} + 4(D+1) H^2 a a' \nabla^2 I^3[F'A''']\;, \label{nabla3}
\end{eqnarray}   
where the following identity was used to get the above desired form:
\begin{eqnarray}
a a' H^2 \Vert \Delta \vec{x} \Vert^2 f(y) = -\frac12 (D-1) I[f](y) - \frac{\nabla \cdot \nabla'}{4 a a' H^2}  I^2[f](y)\;.
\end{eqnarray} 
This result is tabulated in Table ~\ref{ct1}. Similar, but much more tedious work should be done to extract derivatives out for pure temporal and temporal spatial mixed derivatives in equation \eqref{3-pt-ABC}. These results are all tabulated in Tables ~\ref{ct2}, ~\ref{ct3}, ~\ref{ct4}, in the form of the following eleven external operators acting on functions of $y$.
\begin{eqnarray}
\alpha & \equiv & (a a')^2 \square^2 \label{alpha} \; , \\ 
\beta_1 & \equiv & (a a')^2  H^2 \square \; , \\
\beta_2 & \equiv & a a' (a^2 + a^{\prime 2}) H^2 \square \; , \\
\gamma_1 & \equiv & (a a')^2 H^4 \; , \\
\gamma_2 & \equiv & a a' (a^2 + a^{\prime 2}) H^4 \; , \\
\gamma_3 & \equiv & a a' (a + a')^2 H^4 = 2 \gamma_1 + \gamma_2 \; , \\
\delta & \equiv & (a^2 + a^{\prime 2}) \nabla^2 \square \; , \\
\epsilon_1 & \equiv & a a' H^2 \nabla^2 \; , \\
\epsilon_2 & \equiv & (a^2 + a^{\prime 2}) H^2 \nabla^2 \; , \\
\epsilon_3 & \equiv & (a + a')^2 H^2 \nabla^2 = 2\epsilon_1 + \epsilon_2 \; , \\
\zeta & \equiv & \nabla^4 \label{zeta} \; .
\end{eqnarray} 
The last step before completing this section is adding various components the 3-point contribution coming from $A(y)$, $B(y)$ and $C(y)$ in equation \eqref{3-pt-ABC}. The result can be written symbolically as
\begin{eqnarray}
\lefteqn{-iM^2_{\rm 3pt}(x;x')
= -\tilde{\kappa}^2 (aa')^{\frac{D}2 -1} \Biggl\{\alpha f_{\alpha}(y) 
+ \beta_1 f_{\beta_1}(y) + \beta_2 f_{\beta_2}(y)
+ \gamma_1 f_{\gamma_1}(y) + \gamma_2 f_{\gamma_2}(y) + \gamma_3 f_{\gamma_3}(y) 
} \nonumber\\
& & \hspace{4.5cm}+ \delta f_{\delta}(y) 
+ \epsilon_1 f_{\epsilon_1}(y) + \epsilon_2 f_{\epsilon_2}(y) + \epsilon_3 f_{\epsilon_3}(y) 
+ \zeta f_{\zeta}(y) \Biggl \}\;. 
\label{3pt-alphazeta}
\end{eqnarray}
Here the functions, on which eleven external operators are acting, are given in the 
Tables~\ref{ta}, ~\ref{tb1}, ~\ref{tb2}, ~\ref{tg1}, ~\ref{tg2}, ~\ref{tg3}, ~\ref{td}, 
~\ref{te1}, ~\ref{te2}, ~\ref{te3}, and ~\ref{tz} in Appendix \ref{app2}.
This self-mass-squared will eventually be integrated over $d^4x'$ in the effective field equation, \eqref{linear_eq}. Thus, after extracting the derivative operators outside the integral we only need to retain 
$D$ dimension for terms which diverge logarithmically\footnote{Note that $y(x;x')$ vanishes like $(x-x')^2$ at coincidence and so $\int d^4x' 1/y^2$ diverges logarithmically.} and higher at $x=x'$ for $D=4$ in the coefficient functions, $f_{i}(y)$. 

\section{Renormalization}\label{renormalization}

\begin{figure}[htbp]
\centering
\includegraphics[height=0.6cm]{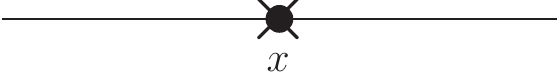}
\caption{Contribution from counterterms.}
\label{cntr}
\end{figure}

In this section we renormalize the scalar self-mass-squared by subtracting counterterms depicted in Fig. \ref{cntr}.
First, we construct counterterms applying the Bogoliubov, Parasiuk, Hepp and Zimmermann (BPHZ) scheme \cite{bogo}.
The structure of our Lagrangian of a scalar, conformally coupled to gravity, allows us to determine three de Sitter invariant counterterms at one loop order. On the other hand, our gauge fixing condition \eqref{gf} breaks de Sitter symmetry, which results in possibility of having de Sitter noninvariant counterterms. It turns out that there is only one noninvariant counterterm because the de Sitter breaking occurs in a particular way.
Identifying these possible counterterms is of great importance for checking 
correctness of the calculation.
Hundreds of terms arising from various places should all add up to terms which respect the symmetries 
not broken due to the gauge fixing term. 
We will soon show this occurs in a highly nontrivial way. 
The next step is to collect all the divergences occurred in the previous section and localize them (in the form of $\delta$-function) so as to be absorbed by the local counterterms. 
Finally, we obtain a finite result which can be used to solve the effective field equation \eqref{linear_eq} at one loop order.
This procedure is summarized in the following two subsections. 

\subsection{Construction of counterterms}

To construct possible counterterms we first note that the superficial degree of divergence for the scalar-graviton interaction in our Lagrangian \eqref{Lagrangian}  at one loop order is four.
This means that in order to cancel those divergences, the corresponding counterterms should have a mass dimension of four. There are also two basic requirements for our case. 
One is, the counterterms must carry two scalar fields, each of which counts one dimension of mass. 
The other is, they should also contain one factor of the loop counting parameter $\kappa^2$ which has the dimension of $mass^{-2}$.
We therefore require each counterterm to have an additional mass dimension of four.
Because our scalar is massless, this can only be achieved by carrying four derivatives.
There are three ways to form an invariant satisfying these requirements.
\begin{itemize}
\item{All four derivatives act on scalars.}
\item{Two derivatives act on scalars and the other two act on the metric.}
\item{All four derivatives act on the metric.}
\end{itemize}
Applying these three ways, we find five invariants listed as follows:
\bea
\kappa^2 \varphi_{;\mu\nu} \varphi_{;\rho\sigma} g^{\mu\nu} g^{\rho\sigma} \sqrt{-g}\;,\; 
\kappa^2 \partial_{\mu} \varphi \partial_{\nu} \varphi R g^{\mu\nu} \sqrt{-g}\;,\;
\kappa^2 \partial_{\mu} \varphi \partial_{\nu} \varphi R^{\mu\nu} \sqrt{-g}\;, \;
\kappa^2 \varphi^2 R^2 \sqrt{-g}\; \mbox{ and }\;  
\kappa^2 \varphi^2 R^{\mu\nu} R_{\mu\nu} \sqrt{-g} \;. 
\label{fourds}
\eea
Finally, specializing to the de Sitter background, i.e., $\hat{R}_{\mu\nu} = (D-1)H^2\hat{g}_{\mu\nu}$,
allows us to put the second and third of \eqref{fourds} together and the fourth and fifth together into one term. This results in three invariant counterterms using the rescaled field, 
$\tilde{\phi}\equiv\phi=a\varphi$, see Eq. \eqref{rescaling_notation},
\bea\Delta \mathcal{L}_1 = \frac12 c_1 \kappa^2 \square \phi \square \phi a^2 \;,\quad
\Delta \mathcal{L}_2 = -\frac12 c_2 \kappa^2 H^2 \partial_{\mu} \phi \partial^{\mu} \phi\;, \quad\Delta \mathcal{L}_3 = \frac12 c_3 \kappa^2 H^4  \phi^2 a^2\;. \label{invar}
\eea
The vertices associated to these counterterms are,
\bea
\frac{i \delta \Delta S_1}{\delta \phi(x) \delta \phi(x')}
\Biggl\vert_{\phi = 0} &=&
i c_1 \kappa^2 a^2 \square^2 \delta^D (x-x'), \label{alp1}
\\
\frac{i \delta \Delta S_2}{\delta \phi(x) \delta \phi(x')}
\Biggl\vert_{\phi = 0} &=&
i c_2 \kappa^2 H^2 a^2 \square \delta^D (x-x'), \label{alp2} 
\\
\frac{i \delta \Delta S_3}{\delta \phi(x) \delta \phi(x')}
\Biggl\vert_{\phi = 0} &=&
i c_3 \kappa^2 H^4  a^2 \delta^D (x-x'). \label{alp3} 
\eea
Here the coefficients $c_i$ will be determined by requirement of canceling the divergences. 

The de Sitter noninvariant counterterm drawn to attention in the beginning of this section
was constructed in Ref. \cite{kahya} by  carefully examining which symmetries are broken by 
our gauge choice \eqref{gf}. Here we review essential points of 
the construction strategy taken by \cite{kahya} and 
finally give the unique noninvariant counterterm. 
The first point is that our gauge fixing term breaks only spatial special conformal transformations
among the full $\frac{1}{2}D(D+1)$ de Sitter isometries:
\begin{itemize}
 \item{$(D-1)$ spatial translations :}
\begin{equation}
 \eta' = \eta\;, ~~ x'^{i} = x^{i} + \epsilon^{i} \;. \label{homog}
\end{equation}
 \item{$\frac{1}{2}(D-1)(D-2)$ rotations :}
\begin{equation}
 \eta' = \eta\;, ~~ x'^{i} = R^{ij}x^{j}\;. \label{isox}
\end{equation}
 \item{$1$ dilation :}
\begin{equation}
 \eta' = k\eta\;, ~~ x'^{i} = kx^{j}\;. \label{dilx}
\end{equation}
 \item{$(D-1)$ spatial special conformal transformations :}
\begin{equation}
 \eta' = \frac{\eta}{1 - 2\vec{\theta}\cdot\vec{x}
+ \parallel\! \vec{\theta} \!\parallel^2 x\cdot x}\;, ~~ x'^{i} =
\frac{x^{i} - \theta^{i}x\cdot x}{1 - 2\vec{\theta}\cdot\vec{x} +
\parallel\! \vec{\theta} \!\parallel^2 x\cdot x}\;. \label{sscx}
\end{equation}
\end{itemize}
Hence we can use the residual symmetries respected by our gauge condition to restrict
the form of noninvariant counterterms. Homogeneity \eqref{homog} requires that 
the counterterm cannot depend on the spatial coordinates $x^i$. Isotropy \eqref{isox} implies
that spatial derivative operators $\partial_i$ must be contracted into each another. 
Dilation symmetry \eqref{dilx} restricts that derivative operators and the conformal time $\eta$ can only appear in the form $a^{-1} \partial_{\mu}$.   
These constraints and a number of partial integration lead us to three possible noninvariant counterterms, 
\be
\frac12 \kappa^2 a^{D-2} \square \varphi \nabla^2 \varphi \;\; , \;\;
\frac12 \kappa^2 a^{D-4} \nabla^2 \varphi \nabla^2 \varphi \;\; {\rm and} 
\;\; -\frac12 \kappa^2 H^2 a^{D-2} \nabla \varphi \cdot \nabla \varphi \; . 
\label{noninv}
\ee
Finally we note that our gauge fixing term \eqref{gf} becomes Poincar\'e invariant in
the flat space limit of $H \rightarrow 0$, where the conformal time $\eta = -e^{-Ht}/H$ with the comoving time $t$ held fixed. Only the final term of \eqref{noninv} vanishes in the flat space limit and it can serve as a noninvariant counterterm. Using the rescaled field it becomes
\bea\Delta \mathcal{L}_4 = -\frac12 \kappa^2 H^2 
\nabla \phi \cdot \nabla \phi  \;,\quad
\eea
with the associated vertex,
\begin{equation}
\frac{i \delta \Delta S_4}{\delta \phi(x) \delta \phi(x')}
\Biggl\vert_{\phi = 0}
= i c_4 \kappa^2 H^2
\nabla^2 \delta^D(x\!-\!x') \; . 
\label{alp4}
\end{equation}

To summarize, there are only four - three invariant and one noninvariant - counterterms allowed for the CC scalar self-mass-squared calculation at one loop order. Any occurrences of divergent terms which do not fit in one of these four counterterms would immediately imply errors in the calculation. Again we emphasize that this provides with a crucial check for perturbative quantum gravity computations like our current work. It is amazing to observe that divergences of the ``non-permitted'' form occurring in various places add up to zero in the first column of TABLE \ref{t3}. Note that the second column becomes finite for $D=4$ by canceling the overall divergent factor $1/(D-4)$.

\subsection{Localization of divergences and a finite result}

Now with the prepared counterterms, it is time to collect all the divergent terms 
and segregate them into a local function in the form of the counterterm vertices, 
that is, an operator acting on the $\delta^D(x-x')$.    
Recall that the contributions from 4-point interactions are already finite and 
the divergences from 3-point vertices involve powers of $y$ that are not integrable for 
$D = 4$ dimensions,
\begin{equation}
\Bigl(\frac{4}{y}\Bigr)^{D}\;, \; 
\Bigl(\frac{4}{y}\Bigr)^{D-1} \quad \mbox{and} \quad \Bigl(\frac{4}{y}\Bigr)^{D-2}\;.
\end{equation}
Again we use the technique of extracting derivative operators to make the terms less singular. 
Specifically we use the following identity to extract d' Alembertian operators from these terms until they
become integrable,
\begin{eqnarray}
\square f(y) = H^2 \Bigl[(4y - y^2) f''(y) + D (2 - y)f'(y)\Bigr]
+{\rm Res} \Bigl[y^{\frac{D}2 -2} f\Bigr] 
\frac{4 \pi^{D/2} H^{2-D}}{\Gamma(\frac{D}2 - 1)} \frac{i}{\sqrt{-g}} \delta^D (x-x').\label{key}
\end{eqnarray}
Here ${\rm Res}[F]$ means the residue of $F(y)$. 
Applying this identity (\ref{key}) each of the nonintegrable terms becomes,
\begin{eqnarray}
\Bigl(\frac{4}{y}\Bigr)^{D}\phantom{\mu\nu}
&=&\frac2{(D-1)D} \frac{\square}{H^2} \Bigl(\frac{4}{y}\Bigr)^{D-1}\;,
\label{0th}\\
\Bigl(\frac{4}{y}\Bigr)^{D-1}
&=&\frac2{(D-2)^2} \frac{\square}{H^2} \Bigl(\frac{4}{y}\Bigr)^{D-2}-\frac2{D-2} \Bigl(\frac{4}{y}\Bigr)^{D-2}\;,
\label{1st}\\
\Bigl(\frac{4}{y}\Bigr)^{D-2}&=&\frac2{(D-4)(D-3)} \frac{\square}{ H^2} \Bigl(\frac{4}{y}\Bigr)^{D-3}-\frac4{D-4} \Bigl(\frac{4}{y}\Bigr)^{D-3}\;.
\label{2nd} 
\end{eqnarray}
Note that the logarithmically divergent power $1/y^{D-2}$ in \eqref{2nd} 
has reduced to the power $1/y^{D-3}$ which is integrable, however it has 
a divergent factor of $1/(D-4)$.
Applying the key identity \eqref{key} to the power of  $1/y^{D/2-1}$ 
we have zero in the form,
\begin{eqnarray}
0=\frac{\square}{H^2} \Bigl(\frac{4}{y}\Bigr)^{\frac{D}2 -1} 
- \frac{D}2 \Bigl(\frac{D}2 -1 \Bigr) \Bigl(\frac{4}{y}\Bigr)^{\frac{D}2 -1} 
-\frac{(4\pi)^{D/2} H^{-D}}{\Gamma(\frac{D}2 -1)} \frac{i}{a^D} \delta^D (x-x') \;. 
\label{delt}
\end{eqnarray}
Adding this expression of zero \eqref{delt} to (\ref{2nd}) allows us to 
segregate the divergence on the local term,
\begin{eqnarray}
\Bigl(\frac{4}{y}\Bigr)^{D-2}
&\!\!\!=\!\!\!&\frac{2}{(D-4)(D-3)}\Biggl\{ 
\frac{(4\pi)^{\frac{D}2} H^{-D}}{\Gamma(\frac{D}2 - 1)}\frac{i \delta^D( x-x')}{a^D} 
+\frac{\square}{H^2} \Biggl[\Bigl(\frac{4}{y} \Bigr)^{D-3} 
- \Bigl(\frac{4}{y}\Bigr)^{\frac{D}2 - 1} \Biggr] \Biggr\} 
\nonumber \\
& & \hspace{3cm} 
-\frac{4}{D-4} \Biggl\{\Bigl(\frac{4}{y}\Bigr)^{D-3} 
- \frac{D (D-2)}{8 (D-3)} \Bigl(\frac{4}{y}\Bigr)^{\frac{D}2 -1} \Biggr\}\;,
\\
&\!\!\!=\!\!\!&\frac{i H^{-D} (4\pi)^{\frac{D}2}}{(D-4)(D-3) \Gamma (\frac{D}2)} (D-2) \frac{\delta^D (x-x')}{a^D} 
-\frac{\square}{H^2} \Biggl\{\frac{4}{y} \ln\Bigl(\frac{y}4\Bigr) \Biggr\} + 2 \Bigl(\frac{4}{y}\Bigr) \ln\Bigl(\frac{y}{4}\Bigr) - \Bigl(\frac{4}{y} \Bigr) + \mathcal{O}(D-4)
\;,\label{D-2} 
\end{eqnarray}
or
\begin{eqnarray}
\lefteqn{
\Bigl(\frac{1}{y}\Bigr)^{D-2}
=\frac{i H^{-D} \pi^{\frac{D}{2}}}{(D-4)(D-3) \Gamma (\frac{D}2) 4^{\frac{D}{2}}} 16(D-2) \frac{\delta^D (x-x')}{a^D}} \nonumber \\ 
& & \hspace{2cm} -\frac{16}{4^D}\frac{\square}{H^2} \Biggl\{\frac{4}{y} \ln\Bigl(\frac{y}4\Bigr) \Biggr\}  
+\frac{32}{4^D} \Bigl(\frac{4}{y}\Bigr) \ln\Bigl(\frac{y}{4}\Bigr)
-\frac{16}{4^D}\Bigl(\frac{4}{y} \Bigr) + \mathcal{O}(D-4)\;.
\label{yup(-2+D)}
\end{eqnarray}
Substituting \eqref{D-2} into \eqref{1st} the quadratically divergent power becomes,
\begin{eqnarray}
\lefteqn{
\Bigl(\frac{1}{y}\Bigr)^{D-1}
= \frac{i H^{-D} \pi^{\frac{D}{2}} }{(D-4)(D-3) \Gamma (\frac{D}{2}) 4^{\frac{D}{2}} } 
\Biggl\{ \frac{8}{D-2} \frac{\square}{H^2} - 8 \Biggr\}\frac{\delta^D (x-x')}{a^D} 
-\frac{2}{4^D} \frac{\square^2}{H^4} \Biggl\{\frac{4}{y} \ln \Bigl(\frac{y}{4}\Bigr) \Biggr\}
}
\nonumber\\
& & \hspace{1.5cm}
+\frac{\square}{H^2} \Biggl\{ \frac{8}{4^D}\Bigl(\frac{4}{y}\Bigr) \ln\Bigl(\frac{y}{4}\Bigr) 
-\frac{2}{4^D} \Bigl(\frac{4}{y} \Bigr)\Biggr\} 
-\frac{8}{4^D} \Bigl(\frac{4}{y} \Bigr)\ln\Bigl(\frac{y}{4}\Bigr) 
+ \frac{4}{4^D} \Bigl(\frac{4}{y} \Bigr) + \mathcal{O}(D-4)\;.
\label{yup(-1+D)}
\end{eqnarray}
Similarly, the quartically divergent term becomes,
\begin{eqnarray}
\lefteqn{
\Bigl(\frac{1}{y}\Bigr)^{D}=\frac{i H^{-D} \pi^{\frac{D}2}}{(D-4) (D-3) \Gamma (\frac{D}2) 4^{\frac{D}2}}
\Biggl\{ \frac4{(D-2)(D-1)D} \frac{\square^2}{H^2} -\frac{4}{(D-1) D} \frac{\square}{H^2}\Biggr\} 
\frac{\delta^D (x-x')}{a^D}
}
\nonumber \\ 
& & \hspace{1cm} -\frac{1}{4^D}\Biggl\{ 
\frac{1}{12} \frac{\square^3}{H^6} \biggl[\frac{4}{y} \ln \Bigl(\frac{y}4 \Bigr) \bigg] 
+\frac{\square^{2}}{H^4} \biggl[\frac13 \Bigl( \frac{4}{y}\Bigr) \ln \Bigl(\frac{y}{4} \Bigr) 
-\frac1{12} \Bigl(\frac{4}{y} \Bigr)\biggr]  
-\frac1{3} \Bigl( \frac{4}{y}\Bigr) \ln\Bigl( \frac{y}{4}\Bigr) + \frac1{6} \Bigl( \frac{4}{y}\Bigr)
\Biggr\} + \mathcal{O}(D-4)\;.
\label{yup(-D)}
\end{eqnarray}
Plugging these expressions into \eqref{3pt-alphazeta} would make the divergent pieces almost look like
the local counterterms except d'Alembertians acting on both powers of $a$ and $\delta^D(x-x')$. 
We employ the following identities to pass all factors of $a$ to the left. 
\begin{eqnarray}
\square a^n \delta^D (x-x')
&=&\Bigl( a^n \square - [n^2 + n(D-1)] H^2 a^n  
-2 n H a^{n-1} \partial_{0} \Bigr) \delta^D (x-x')\;,
\label{ident1}
\\ 
\square ^2 a^n \delta^D (x-x')
&=&\Bigl(a^n \square ^2 - 2 n [3 n+(D-1)] H^2 a^n \square 
-4 n H a^{n-1} \partial_{0} \square  
+ 4 n^3 H^3 a^{n-1} \partial_{0} 
\nonumber \\
& & \hspace{0.5cm} +4 n(n-1) H^2 a^{n-2} \nabla^2 
+n^2 [n+(D-1)]^2 H^4 a^n \Bigr) \delta^D (x-x')\;.
\label{ident2} 
\end{eqnarray}
The following example for the term with the external operator $\alpha$ shows 
how we use these identities in order to put the divergent pieces in the form of the counterterms
\begin{eqnarray}\label{alphay^(2-D)}
& &(aa')^{\frac{D}{2}-1} \alpha \Biggl\{-\frac{(D+2)(D+4) H^{2D-4}\Gamma(\frac{D}2 -1)^2}{128(D-2)(D+1)\pi^D} \Bigl( \frac1{y}\Bigr)^{D-2} \Biggr\} \nonumber \\ 
& & \hspace{0.7cm} =-\frac{(D+2)(D+4) H^{2D-4}\Gamma(\frac{D}2 -1)^2}{128(D-2)(D+1)\pi^D} (aa')^{\frac{D}{2}+1} \square^{2} \Bigl(\frac1{y}\Bigr)^{D-2}\;,\\
& & \hspace{0.7cm} = -\frac{(D+2)(D+4) H^{2D-4}\Gamma(\frac{D}2 -1)^2}{128(D-2)(D+1)\pi^D}\; \frac{i H^{-D} \pi^\frac{D}2}{(D-4)(D-3) \Gamma(\frac{D}2)4^{\frac{D}2}} a^{\frac{D}2+1} 16(D-2) \square ^{2} a^{\frac{D}2+1} \frac{\delta^D (x-x')}{a^D}\;,\\
& & \hspace{0.7cm} = -\frac{(D+2)(D+4)}{2^{5}(D-2)^{3}(D+1)}\; \frac{i H^{D-4}}{(4\pi)^\frac{D}2} \frac{\Gamma(\frac{D}2)}{(D-4)(D-3)} \; \Biggl\{16(D-2) a^2 \square ^2 -8(D-2)^2 (D-4) H^2 a^2 \square \nonumber \\
& &\hspace{1.5cm} + 32(D-2)^2 H a \partial_{0} \square  -8(D-2)^4 H^3 a \partial_{0} + 16(D-2) D H^2 \nabla^ 2- (D-2)^2 D^2 H^4 a^2 \Biggr\} \delta^D (x-x')\;.
\end{eqnarray}
Here $\alpha$ is given in \eqref{alpha}.
Note that the first, second, fifth and sixth terms correspond to 
the counterterm vertices 
\eqref{alp1}, \eqref{alp2}, \eqref{alp4}, and \eqref{alp3}, respectively.  

Using the same procedure for the terms with the remaining ten external operators, 
we can segregate all divergent terms into the form which derivatives acting on $\delta^D(x-x')$. 
The results are given in TABLE~\ref{t1}, TABLE~\ref{t2} 
and TABLE~\ref{t3}.
Note that we have another consistency check for the calculation: 
the contribution to the counterterm \eqref{alp1} vanishes as it must, otherwise this counterterm would not be zero in flat space and break Poincar\'e invariance. 
\begin{table}[htbp]
\begin{center}
\caption{Contributions to counterterms. All terms are multiplied by $\frac{i \tilde{\kappa}^2 H^{D-4}}{(4\pi)^{D/2}} \frac{\Gamma(\frac{D}2)}{
(D-4)(D-3)}$}
\begin{ruledtabular}
\begin{tabular}{lcl}
${\rm External\; operators}$&&${\rm Coef.\;of\;}H^{4}a^{2}\delta^D(x-x')$ \\[1ex]
\hline\\
$\alpha$&&$ \frac{D^2(D+2)(D+4)}{32(D-2)(D+1)}$ \\[2ex]
$\beta_1$&&$ \frac{(D-1)D(D+2)(D+4)}{8(D+1)}$ \\[2ex]
$\beta_2$&&$ -\frac{(D+2)(D+4)(D^2-2D-4)}{8(D+1)}$ \\[2ex]
$\gamma_1$&&$ \frac{D (D^5-35 D^4+67 D^3-2 D^2-264 D-64)}{32(D-2)(D+1)}$ \\[2ex]
$\gamma_2$&&$ \frac{2 D^6+35 D^5-108 D^4-144 D^3+384 D+704}{8(D-2)^{2}(D+1)}$ \\[2ex]
\hline\\
${\rm Total}$&&$\frac{D^7-29 D^6+282 D^5-540 D^4-856 D^3+224 D^2+1664 D+3328}{32(D-2)^{2}(D+1)}$ \\[1ex]
\end{tabular}\label{t1}
\end{ruledtabular}
\end{center}
\end{table}

\begin{table}[htbp] 
\begin{center}
\caption{Contributions to counterterms. All terms are multiplied by $\frac{i \tilde{\kappa}^2 H^{D-4}}{(4\pi)^{D/2}}
\frac{\Gamma(\frac{D}2)}{(D-4)(D-3)}$.}
\begin{ruledtabular}
\begin{tabular}{lccc}
${\rm External}$&${\rm Coef.\;of}$&${\rm Coef.\;of}$&${\rm Coef.\;of}$\\[1ex]
${\rm operators}$&$a^{2}\square^2\delta^D(x-x')$&$H^{2}a^{2}\square\delta^D(x-x')$&$H^{2}\nabla^2 \delta^D(x-x')$\\[1ex]
\hline\\
$\alpha$&$-\frac{(D+2)(D+4)}{2(D-2)^{2}(D+1)}$&
$\frac{(D-4)(D+2)(D+4)}{4(D-2)(D+1)}$&
$-\frac{D(D+2)(D+4)}{2(D-2)(D+1)}$ \\[2ex]
$\beta_1$&$0$&$\frac{(D-1)(D+2)(D+4)}{2(D-2)(D+1)}$&$0$ \\[2ex]
$\beta_2$&$0$&$-\frac{(D+2)(D+4)}{2(D+1)}$&$0$ \\[2ex]
$\gamma_1$&$\frac{(D-20)(D+2)}{2(D-2)^2 (D+1)}$&
$\frac{D^5-37 D^4+116 D^3-172 D^2-496 D+576}{8(D-2)^2 (D+1)}$&
$\frac{(D-20)D(D+2)}{2(D-2)(D+1)}$ \\[2ex]
$\gamma_2$&
$\frac{12(D+2)}{(D-2)^2 (D+1)}$&
$\frac{D^4+12 D^3-5 D^2-40 D-184}{(D-2)^2(D+1)}$&
$\frac{12(D^3+8)}{(D-2)^{2}(D+1)}$ \\[2ex]
$\epsilon_1$&$0$&$0$&$\frac{D^4+12 D^3-24 D^2-20 D+16}{(D-2)(D^{2}-1)}$ \\[2ex]
$\epsilon_2$&$0$&$0$&$-\frac{13D^{2}+6D-16}{(D^{2}-1)}$ \\[2ex]
$\epsilon_3$&$0$&$0$&$\frac{2(D-2)(D+2)(D+4)}{(D^{2}-1)}$ \\[2ex]
\hline\\
${\rm Total}$&$0$&$\frac{(D-4)(D^4-23 D^3+124 D^2+260 D+208)}{8(D-2)^2 (D+1)}$&
$\frac{3 D^5-3 D^4-34 D^3+96 D^2+112 D-192}{(D-2)^{2}(D^{2}-1)}$ \\[1ex]
\end{tabular}\label{t2}
\end{ruledtabular}
\end{center}
\end{table}
\begin{table}[htbp]
\begin{center}
\caption{Other contributions to counterterms. All terms are multiplied by $\frac{i \tilde{\kappa}^2 H^{D-4}}{(4\pi)^{D/2}}
\frac{\Gamma(\frac{D}2)}{(D-4)(D-3)}$.}
\begin{ruledtabular}
\begin{tabular}{lccc}
${\rm External}$&${\rm Coef.\;of}$ &${\rm Coef.\;of}$ \\[1ex]
${\rm operators}$&$H a \partial_{0}\square \delta^D(x-x')$&$H^3 a \partial_{0}\delta^D(x-x')$ \\[1ex]
\hline\\
$\alpha$&$-\frac{(D+2)(D+4)}{(D-2)(D+1)}$&$\frac{(D-2)(D+2)(D+4)}{4(D+1)}$ \\[2ex]
$\beta_1$&$0$&$\frac{(D-1)(D+2)(D+4)}{2(D+1)}$ \\[2ex]
$\beta_2$&$0$&$-\frac{(D-2)(D+2)(D+4)}{2(D+1)}$ \\[2ex]
$\gamma_1$&$\frac{(D-20)(D+2)}{(D-2)(D+1)}$&
$\frac{D^5-37 D^4+112 D^3-92 D^2-480 D+256}{8(D-2)(D+1)}$ \\[2ex]
$\gamma_2$&$\frac{24(D+2)}{(D-2)(D+1)}$&
$\frac{D^4+12 D^3-5 D^2-64 D-232}{(D-2)(D+1)}$ \\[2ex]
\hline\\
${\rm Total}$&$0$&$\frac{(D-4)(D^4-23 D^3+124 D^2+356 D+400)}{8(D-2)(D+1)}$ \\[1ex]
\end{tabular}\label{t3}
\end{ruledtabular}
\end{center}
\end{table}

Besides the local divergent terms, the remaining pieces are nonlocal and finite for $D=4$ dimension. 
Again we illustrate how the nonlocal finite terms are identified using the case with the $\alpha$ operator.
\begin{eqnarray}\label{finitealphay^(2-D)}
& &(aa')^{\frac{D}{2}-1} \alpha \Biggl\{-\frac{(D+2)(D+4) H^{2D-4}\Gamma(\frac{D}2 -1)^2}{128(D-2)(D+1)\pi^D} \Bigl( \frac1{y}\Bigr)^{D-2} - \frac{3H^4}{32\pi^4}\frac1{y} \Biggr\} \nonumber \\ 
& & \hspace{0.7cm} =-\frac{(D+2)(D+4) H^{2D-4}\Gamma(\frac{D}2 -1)^2}{128(D-2)(D+1)\pi^D} (aa')^{\frac{D}{2}+1} \square^{2} \Bigl( \frac1{y} \Bigr)^{D-2}- (aa')^{\frac{D}2+1} \square^{2} \frac{3H^4}{32\pi^4}\frac1{y}\;,\\
& & \hspace{0.7cm} = -\frac{(D+2)(D+4) H^{2D-4}\Gamma(\frac{D}2 -1)^2}{128(D-2)(D+1)\pi^D}\; (aa')^{\frac{D}2+1} \square^{2} \Biggl\{ -\frac{16}{4^D}\frac{\square}{H^2} \Biggl\{\frac{4}{y} \ln\Bigl(\frac{y}4\Bigr) \Biggr\} +\frac{32}{4^D} \Bigl(\frac{4}{y}\Bigr) \ln\Bigl(\frac{y}{4}\Bigr) -\frac{16}{4^D}\Bigl(\frac{4}{y} \Bigr) \Biggr\}\;,\nonumber\\
& & \hspace{2cm} -(aa')^{\frac{D}2+1} \square^{2} \frac{3H^4}{32\pi^4}\Bigr(\frac1{y}\Bigl) \;,\\
& & \hspace{0.7cm} = -\frac3{80} \frac{H^4}{\pi^{4}} (aa')^3 \square^{2} \Biggr\{ -\frac{16}{4^4} \frac{\square}{H^2} \Bigl( \frac{\ln{x}}{x} \Bigr) + \frac{32}{4^4} \Bigl(\frac{\ln{x}}{x} \Bigr) - \frac{16}{4^4} \Bigl(\frac1{x} \Bigr) \Biggl\} - (aa')^3 \square^2 \frac{3H^4}{32\pi^4} \Bigl(\frac1{4x}\Bigr)\;,\nonumber\\
& & \hspace{0.7cm} = \frac3{1280} \frac{H^2}{\pi^{4}} (aa')^3 \square^{3} \Bigl(\frac{\ln{x}}{x} \Bigr) -\frac3{640} \frac{H^4}{\pi^{4}} (aa')^3 \square^2 \Bigl( \frac{\ln{x}}{x}+\frac9{2x}\Bigr)\;.
\end{eqnarray}
This time we only write the nonlocal finite pieces without $\delta^D(x-x')$ from 
the power $(\frac1{y})^{D-2}$ in \eqref{yup(-2+D)}.
In the final two lines, we re-define the function $y(x;x')$ in \eqref{ydef} as $x \equiv \frac{y}{4}$. 
Also, note that we take the $D=4$ limit to get the final result. 
Using the same method for the terms with the other external operators 
we obtain nonlocal finite terms for each case, i.e., each external operator.
These newly found finite terms are added to the already found finite terms listed 
in TABLES ~\ref{ta} - ~\ref{tz}. The summation of all the finite nonlocal terms 
from the 3-point interactions are given in TABLE ~\ref{t4}. 

\begin{table}[htbp]
\begin{center}
\caption{All Finite Nonlocal Contributions with $x \equiv \frac{y}{4}$, where $y(x;x')$ is defined in the equation (\ref{ydef}).}
\begin{ruledtabular}
\begin{tabular*}{1.5\textwidth}{ll}
$\phantom{ssssssssss}\rm External \; operators$&$\rm Coef. \; of \; \frac{\tilde{\kappa}^2 H^4}{(4\pi)^4}$\\[1ex]
\hline\\
$\phantom{ssssssssss}(aa')^3 \square^3/H^2$&$\frac{9\ln{x}}{5x}\phantom{ssssssssssssssssssssss}$\\[2ex]
$\phantom{ssssssssss}(aa')^3 \square^2$&$\frac{9\ln{x}}{x} - \frac{21}{5x}$\\[2ex]
$\phantom{ssssssssss}(aa')^3 H^2\square$&$-\frac{267\ln{x}}{5x} + \frac{51}{x}$\\[2ex]
$\phantom{ssssssssss}(aa')^3 H^4$&$\frac{258\ln{x}}{5x} - \frac{549}{5x}$\\[2ex]
$\phantom{ssssssssss}(aa')^2(a^2+a'^2)\square^3/H^2$&$-\frac{9\ln{x}}{10x}$\\[2ex]
$\phantom{ssssssssss}(aa')^2(a^2+a'^2)\square^2$&$-\frac{99\ln{x}}{10x} - \frac{9}{10x}$\\[2ex]
$\phantom{ssssssssss}(aa')^2(a^2+a'^2)H^2\square$&$\frac{36\ln{x}}{x} - \frac{51}{2x}$\\[2ex]
$\phantom{ssssssssss}(aa')^2(a^2+a'^2)H^4$&$-\frac{108\ln{x}}{5x} + \frac{174}{5x}$\\[2ex]
$\phantom{ssssssssss}(aa')^2 H^2 \nabla^2$&$\frac{128\ln{x}}{5x} + \frac{166}{5x}$\\[2ex]
$\phantom{ssssssssss}(aa')^2 \nabla^2 \square$&$-\frac{64\ln{x}}{5x}$\\[2ex]
$\phantom{ssssssssss}(aa')(a^2+a'^2) H^2 \nabla^2$&$-\frac{4\ln{x}}{x} - \frac{2}{x}$\\[2ex]
$\phantom{ssssssssss}(aa')(a^2+a'^2)\nabla^2\square$&$\frac{2\ln{x}}{x} + \frac{2}{x}$\\[2ex]
$\phantom{ssssssssss}(aa')\nabla^4$&$0$\\[2ex]
\end{tabular*}\label{t4}
\end{ruledtabular}
\end{center}
\end{table}

Our final result for the regulated self-mass-squared consists of three finite parts: 
\begin{itemize}
\item{the local 4-point contributions given in \eqref{4pt-4dim},}
\item{the local 3-point contributions coming from TABLES ~\ref{t1} - ~\ref{t3},}
\item{the nonlocal pieces coming from 3-point interaction, i.e. TABLE ~\ref{t4},}
\end{itemize}
It has the following form, 
\begin{eqnarray}
\lefteqn{-iM^2_{\rm reg}(x;x') = -i \kappa^2 a^2 \Bigl( d_1 \square^2 + d_2 H^2 \square + d_3 H^4 + d_4 H^2 \frac{\nabla^2}{a^2} \Bigr) \delta^D(x-x')}
\nonumber \\
&& + {\rm Table ~\ref{t4}} + 2 i \kappa^2 \times \frac{H^{4}a^2}{16\pi^2} \delta^4 (x-x') + \frac{9 i \kappa^2}{40} \times \frac{H^{3}a}{16\pi^2} \partial_0 \delta^4 (x-x') + \mathcal{O}(D-4)\;.
\label{Total reg}
\end{eqnarray}
The last term in the above equation comes from the second column of Table ~\ref{t3}. Here the coefficients $d_i$ are 
\begin{eqnarray}
d_1 &=& 0\;, \\
d_2 &=& \frac{H^{D-4}}{(4 \pi)^{\frac{D}{2}}} \biggl[  \frac{7}{80} + \mathcal{O}(D-4) \biggr]\;, \\
d_3 &=& \frac{H^{D-4}}{(4 \pi)^{\frac{D}{2}}} \biggl[ 
\frac{(D^7-29 D^6+282 D^5-540 D^4-856 D^3+224 D^2+1664 D+3328)\Gamma(\frac{D}2)}
{2^{11}(D-4)(D-3)(D-1)^{2}(D+1)} \biggr]\nonumber \;,\\
&=& \frac{H^{D-4}}{(4 \pi)^{\frac{D}{2}}} \biggl[ \frac{3}{40 (D-4)} + \frac{1021}{3600} - \frac{3 \gamma}{80} + \mathcal{O}(D-4) \biggr] \;,\\
d_4 &=& \frac{H^{D-4}}{(4 \pi)^{\frac{D}{2}}} \biggl[ \frac{(3 D^5-3 D^4-34 D^3+96 D^2+112 D-192)\Gamma(\frac{D}2)}{2^{6}(D-4)(D-3)(D-1)^{2}(D^{2}-1)} \biggr]\nonumber \;,\\
&=& \frac{H^{D-4}}{(4 \pi)^{\frac{D}{2}}} \biggl[ \frac{2}{9 (D-4)} - \frac{59}{540} - \frac{\gamma}{9} + \mathcal{O}(D-4) \biggr]\;.
\end{eqnarray}
Here $\gamma$ is Euler's constant which is approximately equal to 0.577215. It appears here due to the expansion of the Gamma function. Following the BPHZ renormalization scheme, we will choose $c_i$'s to absorb the divergent terms:
\begin{equation}
c_i = -d_i + \Delta c_i \end{equation}
where $\Delta c_i$ is the remaining arbitrary finite term for each of the four $c_i$.
Taking the unregulated limit (D=4) provides us with the final renormalized result 
for the conformal-conformal interaction part of the full self-mass-squared at one loop order,
\begin{eqnarray}
\lefteqn{-iM^2_{\rm ren}(x;x') = -i \kappa^2 a^2 \Bigl( \Delta c_1 \square^2 + \Delta c_2 H^2 \square + \Delta c_3 H^4 + \Delta c_4 H^2 \frac{\nabla^2}{a^2} \Bigr) \delta^4(x-x')}
\nonumber \\
&& + {\rm Table ~\ref{t4}} + 2 i \kappa^2 \times \frac{H^{4}a^2}{16\pi^2} \delta^4 (x-x') + \frac{9 i \kappa^2}{40} \times \frac{H^{3}a}{16\pi^2} \partial_0 \delta^4 (x-x') + \mathcal{O}(D-4)\;.
\label{Total ren}
\end{eqnarray}

\section{Discussion}\label{dis}

We have calculated the self-mass-squared of a conformally coupled scalar interacting 
with a graviton at one loop order on a locally de Sitter geometry. 
A technical advantage of the scalar being conformally coupled to gravity is that 
we can do the computations around the flat space metric. Specifically, we can use 
the conformally rescaled scalar propagator for flat space.
On the other hand, the graviton propagator stays the same because it is not conformally invariant.
The computation was done using dimensional regularization and renormalized by 
absorbing divergences with BPHZ counterterms. 
The fully renormalized result is given in Eq. \eqref{Total ren} with the finite nonlocal contributions in TABLE \ref{t4}.

The purpose of this computation is to investigate quantum gravitational effects to the scalar mode function during Inflation.
Despite the fact that our scalar is taken to be a spectator field during inflation
driven by cosmological constant $\Lambda$, one might still apply the result to the inflaton of scalar-driven inflation because its potentials are considered to be nearly flat. 
The first step for checking whether or not gravitons give a significant correction to scalars is to compute and renormalize the scalar self-mass-squared $-iM^2(x;x')$, which concerns our present paper. 
The second step is to solve the quantum corrected effective field equation \eqref{linear_eq}. 
At this point it should be noted that this field equation \eqref{linear_eq} is derived in the in-out formalism which one typically uses for scattering amplitude computations in flat space. However, in a time-dependent background such as de Sitter, the in-vacuum does not evolve to the out-vacuum and thus the in-out formalism is not applicable to our case. Therefore we will instead use the in-in formalism also called the Schwinger-Keldysh formalism \cite{Sch, KTM, LVK, Jrd, CHU, CVE, FW, SW11} in which the effective field equation \eqref{linear_eq} is replaced by
\begin{equation}
a^4 \square \phi(x) - \frac16 R \phi(x) - \int_{\eta_i}^0 \! d\eta'\!\int\!d^3x' \,
\Bigl\{M^2_{\scriptscriptstyle ++}(x;x') + M^2_{\scriptscriptstyle +-}(x;x')
\Bigr\} \phi(x') = 0 \; . \qquad \label{linear_eq_ren}
\end{equation}
Here in order to get $-iM^2_{\scriptscriptstyle ++}(x;x')$ and $-iM^2_{\scriptscriptstyle +-}(x;x')$ (which are composed of $i\Delta_{\scriptscriptstyle ++}(x;x')$ and $i\Delta_{\scriptscriptstyle +-}(x;x')$ respectively), we make the following substitutions for $y(x;x')$,
\begin{eqnarray}
i\Delta_{\scriptscriptstyle ++}(x;x')\, : & & y \rightarrow
y_{\scriptscriptstyle ++}(x;x') \equiv a(\eta) a(\eta')H^2 \Bigl[ \Vert \vec{x}
\!-\! \vec{x}' \Vert^2 - (\vert \eta\!-\! \eta'\vert \!-\! i \delta )^2 \Bigr]
\; , \label{y++} \qquad \\
i\Delta_{\scriptscriptstyle +-}(x;x')\, : & & y \rightarrow
y_{\scriptscriptstyle +-}(x;x') \equiv a(\eta) a(\eta')H^2 \Bigl[ \Vert \vec{x}
\!-\! \vec{x}' \Vert^2 - (\eta\!-\! \eta' \!+\! i \delta )^2 \Bigr]
\;. \label{y+-} \qquad
\end{eqnarray}
Note that $-iM^2_{\scriptscriptstyle ++}(x;x')$ is the same as the usual in-out self-mass-squared $-iM^2(x;x')$.

Our aim is to determine if the tree order mode function of the CC scalar (see for example, \cite{W-QGage})
\be
\phi_0(t,\vec{x}) = u(t,k)e^{i\vec{k}\cdot\vec{x}}\;, \quad
\mbox{where} \quad
u(t,k) = \sqrt{\frac{\hbar}{2ck}}\frac{\exp\Big[-ick\int_{t}^{t_i}\frac{dt'}{a(t')}\Big]}{a(t)}\;,
\ee
gets corrections from the self-mass-squared at one loop order, which will be a future paper.

The conformally coupled scalar field interacts with gravity via a kinetic term as well as the conformal coupling term. The case with no conformal coupling, which is called minimal coupling, was studied in the previous work \cite{kahya}. In this paper we examined the other one, which is conformal coupling. For the case of 4-point interactions depicted in Fig. \ref{4pt}, adding these two results will suffice. However, for the case of 3 point interactions graphically represented in Fig. \ref{3pt}, one has the possibility of mixing the two. One vertex can be a kinetic interaction and the other can come from the conformal coupling.
In the current work, we have completed only the half of the first step, namely calculating the one loop scalar self-mass-squared from conformal coupling. In a subsequent paper we will include the purely kinetic terms \cite{kahya} and the mixing of kinetic and conformal interactions in Eq. \eqref{M^2_3point_cross} and use these full results in order to solve the effective field equation  \eqref{linear_eq_ren}.

We also would like to highlight a couple of points to corroborate correctness of our calculation, which is a question that comes into mind for any computation of this length. The first one is that the most divergent part of the graviton propagator (which consists of the A, B and C parts) is same as the conformally coupled scalar propagator as one can see from Eqs. \eqref{DeltaAA}-\eqref{DeltaC}. Adding the tensor factors of the A, B and C parts of the graviton propagator gives
\be 
2 \overline{\eta}_{\mu (\rho} \overline{\eta}_{\sigma) \nu} -
\frac{2\overline{\eta}_{\mu\nu} \overline{\eta}_{\rho \sigma}}{D-3}  -4 \delta^0_{(\mu} \overline{\eta}_{\nu) (\rho} \delta^0_{\sigma)}
+ \frac{2\Big[(D-3) \delta^0_{\mu} \delta^0_{\nu} + \overline{\eta}_{\mu\nu}\Big] 
\Big[(D-3)\delta^{0}_{\rho}\delta^{0}_{\sigma} + \overline{\eta}_{\rho\sigma}\Big]}{(D-3)(D-2)} 
 =  2\eta_{\mu(\rho} \eta_{\sigma)\nu} - \frac{2}{D-2} \eta_{\mu\nu} \eta_{\rho\sigma} \;, 
\ee 
which is the tensor factor of the graviton propagator in flat space. This means that the most singular terms in the full computation are identical to those of flat space. This is why the sum of the coefficients for  $a^2 \Box^2 \delta^{D}(x-x')$ becomes zero in TABLE ~\ref{t2}. The second, and much stronger, check comes from the analysis of the de Sitter noninvariant counterterms. The de Sitter symmetry is broken, due to the form of our gauge fixing term, in such a way only a subgroup of the full de Sitter group is preserved. Respecting the remaining symmetries turns out to make the divergent terms $H a \partial_0  \Box \delta^{D}(x-x')$ and $H^3 a \partial_0 \delta^{D}(x-x')$ become zero. This is exactly what happens in TABLE ~\ref{t3} in a highly nontrivial way. 
And finally the last check is getting zero for $\nabla^4$ in TABLE ~\ref{t4}. Since working with the most singular term is similiar to working in flat space where the gauge fixing term vanishes, one can at most get quadratic divergence. Therefore the de Sitter breaking gauge fixing term can only occur at quadratically divergent level. This will suffice to give only $\nabla^2$ which breaks de Sitter invariance but still respects the other symettries that are not broken. 

One point worth to note is that the result for the minimal coupling only. The massless, minimally coupled scalar gets no significant corrections from inflationary gravitons at one loop order \cite{kahya, kahya2}. That reason why is that, even though inflation produces the vast ensemble of gravitons, they only interact with the MMC scalar via kinetic energy which redshifts to zero at late times. Adding conformal coupling engenders another interaction, which is non-derivative and so not redshifted. 
There is clearly one term that might give a big effect, which is the $H^4 a^2 \delta^4(x-x')$ term in Eq. \eqref{Total ren}. This term was not present in the MMC calculation studied in \cite{kahya, kahya2}. But, unfortunately this can be removed by the conformal counterterm, $a^2 \Delta c_3 H^4 \delta^4(x-x')$ with an arbitrary constant  $\Delta c_3$ in front. However the finite terms we got here may give some enhancement after we solve the effective field equation. In \cite{kahya2} the corresponding finite terms nontrivially added up to be zero but there is no a priori reason why they should add up to zero at the end in this calculation. Also including the MMC-CC cross terms might give an interesting effect and checking it explicitly is our goal for the series of this and subsequent papers. 

Finally the most interesting and cosmologically relevant part would be the second step of solving the quantum corrected effective field equation for the scalar mode function. If the mode function gets quantum corrections, so does the power spectrum. Weinberg's analysis \cite{SW11} suggests that the possibility of quantum loop corrections to the power spectra of primordial density perturbations. Our case, if we find any correction, would serve as a specific example for his analysis. 
In the age of possibility of observing primordial gravitational waves with a detector like 
BICEP2 \cite{BICEP2}, resolving quantum corrections due to gravitons, even though it may still take a few decades more, does not sound utterly hopeless. 
We hope to provide a foreground for those future measurements.

\section{Acknowledgment}\label{ack}
We deeply thank Richard P. Woodard for helpful comments and discussions. 
SP also thanks Ivan Agullo for comments on the scalar 2-point correlators.  
SB and EOK acknowledge support from Tubitak Grant Number:112T817. 
SP is supported by the Eberly Research Funds of the Pennsylvania State University.



\appendix


\section{Tables for Indentities to extract derivatives}
\label{app1}

\begin{table}[htbp]
\caption{For $F(y)\nabla^{4}A(y)$.}
\begin{ruledtabular}
\begin{tabular}{|l|l|}
\hline
$\rm{External}$& \\ 
$\rm{operators}$&$\rm{The\; functions \; they \; act \; on}$\\
\hline \hline
$\gamma_1 $&$4(D^2-1) (I[F'A'']-I^2[F'A^{(3)}])$\\
\hline \hline
$\epsilon_1 $&$4(D+1)I^3[F'A^{(3)}]$\\
\hline \hline$\delta$&$I^4[FA^{(4)}]$\\
 \end{tabular}\label{ct1}
\end{ruledtabular}
\end{table}

\begin{table}[htbp]
\caption{For $F(y)\partial_{0}^{2}\partial_{0}^{'2}A(y)$.}
\begin{ruledtabular}
\begin{tabular}{|l|l|}
\hline
$\rm{External}$&\\ 
$\rm{operators}$&$\rm{The\; functions \; they \; act \; on}$\\
\hline \hline
$\alpha $&$I^4[F A^{(4)}]$\\
\hline\hline
$\beta_1 $&$\frac1{2}(D-2)^{2}I^4[F A^{(4)}]$\\
\hline \hline
$\beta_2 $&$-(D-2)(D-1)I^4[F A^{(4)}]$\\
\hline \hline
&$\frac12 (D-2)^3 (D-1) I^4[F A^{(4)}]-2(D-1)(I^2[F' A'] + I^3[F' A''] + I^4[F' A^{(3)}])$\\
$\gamma_1 $&$+(12+40y-26y^2)(I[F'A''] + I^2[F'A^{(3)}])+8(D-1) (I^2[F'A''] + I^3[F'A^{(3)}])+(-48y+64y^2-20y^3) I[F'A^{(3)}]$\\
&$\;\;+(D-1)(-24 I^2[F'A^{(3)}] + 28 I[y I[F'A^{(3)}]] -14 I[y^2 I[F'A^{(3)}]])-20(D-1) I[y I[F'A''] + y I^2[FA^{(3)}]]$\\
\hline \hline
$\gamma_2 $&$\frac12 (D-2)^2 y I^3[FA^{(4)}]] -32y (I[F'A''] + I^2[F A^{(3)}])- 12(D-1)( I^2[F'A''] + I^3[F'A^{(3)}] )$\\
&$\;\;+ (D-1) (24I^2[F'A^{(3)}]- 12 I[y I[F'A^{(3)}]])+(48y-26y^2) I[F'A^{(3)}]$\\
\hline \hline
$\gamma_3 $&$\frac12 (D-2)^2(y^2I^2[F A^{(4)}] + y I^3[F A^{(4)}]) +\frac12 (D-2)^3 y I^3[F A^{(4)}]$\\
\hline \hline
$\delta $&$- I^4 [ F A^{(4)}](y)$\\
\hline \hline
&$\frac12 (D-2) 2 D  I^4[F A^{(4)}] + 2(D-2) I[y I^2[FA^{(4)}] ] -\frac12 (D-2) I[y^2 I^2[FA^{(4)}] ] -\frac12 (D-2)D I[y I^3[FA^{(4)}]] $\\
$\epsilon_1 $&$-\frac14 (D-2)^3 I^5[F A^{(4)}] -(D-2)^2 I^4[F A^{(4)}] -(D-2)(D-1) I^4[FA^{(4)}] +(I^3[F'A'] + I^4[F'A''] + I^5[F'A^{(3)}])$\\
&$\;+10 I^2[y I[F'A''] + y I^2[F'A^{(3)}] ] -4( I^3[F'A''] + I^4[F'A^{(3)}] ) + 7 I^2 [y^2 I[F'A^{(3)}] ]- 14 I^2 [y I[F'A^{(3)}] ] +12 I^3[F'A^{(3)}]$\\
\hline \hline
$\epsilon_2 $&$6 ( I^3[F'A'']+I^4[F 'A^{(3)}] ) + 6 I^2[y I[F'A^{(3)}]]-12 I^3[F'A^{(3)}]$\\
\hline \hline
$\epsilon_3 $&$-(D-2)y I^3[FA^{(4)}]$\\
\hline \hline
$\zeta $&$I^4[F A^{(4)}] +\frac12 (D-2) I^5[F A^{(4)}])$\\
\end{tabular}\label{ct2}
\end{ruledtabular}
\end{table}

\begin{table}[htbp]
\caption{For $F(y)\nabla^{2}(\partial_{0}^{2}+\partial_{0}^{'2})A(y)$.}
\begin{ruledtabular}
\begin{tabular}{|l|l|}
\hline
$\rm{External}$& \\ 
$\rm{operators}$&$\rm{The\; functions \; they \; act \; on}$\\
\hline \hline
&$ 8(D-1)^2 (I^2[F'A''] + I^3[F'A^{(3)}]) -8(D-1) (I[F'A''] + I^2[F' A^{(3)}])+16(D-1) y (I[F'A''] + I^2[F'A^{(3)}])\;\;$\\
$\gamma_1 $&$+(-16(D-1)y + 8(D-1)y^2) I[F'A^{(3)}] -8(D-1)(D+4) I^2[F'A^{(3)}]$\\
&$\;\;+4(D-1)^2 I[yI[F'A^{(3)}]]$ \\
\hline \hline
$\gamma_2 $&$4(D-1) ( I[F'A'] + I^2[F'A''] + I^3[F'A^{(3)}])- 12(D-1) (I^2[F'A''] + I^3[F'A^{(3)}])$\\
&$\;\;+8(D-1)y (I[F'A''] + I^2[F'A^{(3)}])+2(D-1)y^2 I[F'A^{(3)}]+24(D-1) I^2[F'A^{(3)}]-12(D-1)I[y I[F' A^{(3)}]]$\\
\hline \hline
$\delta  $&$- I^4 [ F A^{(4)}]$\\
\hline \hline
$\epsilon_1 $&$-(D-2)(D-1)I^4[FA^{(4)}] + 4(D-1) (I^3[F'A''] +I^4[F'A^{(3)}])+ 4(D+4) I^3[F'A^{(3)}] -2(D-1) I^2 [y I[F'A^{(3)}]]$\\
\hline \hline
$\epsilon_2 $&$ 6 ( I^3[F'A'']+I^4[F 'A^{(3)}] ) + 6 I^2[y I[F'A^{(3)}]]-12 I^3[F'A^{(3)}] $\\
\hline \hline
$\epsilon_3 $&$-(D-2)y I^3[FA^{(4)}]$\\
\hline \hline
$\zeta $&$2 I^4[FA^{(4)}] +\frac12 (D-2) I^5[F A^{(4)}]$\\
\end{tabular}\label{ct3}
\end{ruledtabular}
\end{table}

\begin{table}[htbp]
\caption{For $F(y)\nabla^{2}(\partial_{0}\partial_{0}^{'2})B(y)$.}
\begin{ruledtabular}
\begin{tabular}{|l|l|}
\hline
$\rm{External}$& \\ 
$\rm{operators}$&$\rm{The\; functions \; they \; act \; on}$\\
\hline \hline
&$2(D-1) ( I[F'B'] + I^2[F'B''] + I^3[F'B^{(3)}] )-4(D^2-1) (I^2[F'B''] + I^3[F'B^{(3)}])$\\
$\gamma_1 $&$\;\;+((4-2y)(D-1)) (I[F'B''] +I^2[F'B^{(3)}])+(8(D-1)y - 2(D-1)y^2) I[F'B^{(3)}]$\\
&$\;\;+4((D-1)(D+4)) I^2[F'B^{(3)}] -2((D-1)(D+4)) I[yI[F' B^{(3)}]]$\\
\hline \hline
$\gamma_2 $&$-12(D-1)  I^2[F'B^{(3)}]$\\
\hline \hline
$\delta  $&$\frac12 I^4 [ F B^{(4)}]$\\
\hline \hline
$\epsilon_1 $&$\frac12 (D-2)(D-1) I^4 [FB^{(4)}]+ 2(D+1) ( I^3[F'B'']+I^4[F'B^{(3)}] )+ (D+4) I^2[y I[F'B^{(3)}]] - 2(D+4) I^3[F'B ^{(3)}]$\\
\hline \hline
$\epsilon_2 $&$ \frac12 y I^3[FB^{(4)}] + 6 I^3[F'B^{(3)}]$\\
\hline \hline
$\epsilon_3 $&$ \frac12 y^2 I^2[FB^{(4)}]+ \frac12 (D-1) y I^3[FB^{(4)}]$\\
\hline \hline
$\zeta  $&$ -I^4[FB^{(4)}] -\frac14 (D-2) I^5[F B^{(4)}]$\\
\end{tabular}\label{ct4}
\end{ruledtabular}
\end{table}

\section{Tables for the coefficient functions of external operators}
\label{app2}

\begin{table}[htbp]
\caption{For $\alpha \equiv (aa')^2 \square^2 $ type terms.}
\begin{ruledtabular}
\begin{tabular}{|l|l|}
\hline
\multicolumn{2}{|c|}{}\\
  \multicolumn{2}{|c|}{$ f_{\alpha}(y) = C_{1a} \;\;f_{\alpha(1a)} (y) + C_{1c} \;\;f_{\alpha(1c)} (y),$}\\
  \multicolumn{2}{|c|}{}\\
\hline \hline
$f_{\alpha(1a)}(y)$&$I^4[F A^{(4)}]\phantom{ssssssssssssssssssssssssssssssssssssssssssssssssssss}$\\
\hline \hline
$f_{\alpha(1c)}(y)$&$I^4[F C^{(4)}]$\\
\hline \hline
\hline
\multicolumn{2}{|c|}{}\\
\multicolumn{2}{|c|}{${\rm Total\;for\;}f_{\alpha}(y)$}\\
\multicolumn{2}{|c|}{}\\
\hline \hline
\multicolumn{2}{|c|}{}\\
\multicolumn{2}{|c|}{$-\frac{(D+2)(D+4)H^{2D-4}\pi^{-D}\Gamma{(\frac{D}2-1})^2}{128(D-2)(D+1)} \; (\frac1{y})^{D-2} - \frac{3H^4}{32\pi^4} \; \frac1{y}.$}\\
\multicolumn{2}{|c|}{}\\
\end{tabular}\label{ta}
\end{ruledtabular}
\end{table}

\begin{table}[htbp]
\caption{For $\beta_1 \equiv (aa')^2 H^2 \square$ type terms.}
\begin{ruledtabular}
\begin{tabular}{|l|l|}
\hline
\multicolumn{2}{|c|}{}\\
\multicolumn{2}{|c|}{$f_{\beta_1}(y) = C_{1a}\;\;f_{\beta_1(1a)} (y)+ C_{1c}\;\;f_{\beta_1(1c)} (y),$}\\
\multicolumn{2}{|c|}{}\\
\hline \hline
$f_{\beta_1(1a)}(y)$&$-(D-2)(D-1)\;\;I^4 [F A^{(4)}]\phantom{sssssssssssssssssssssssssssssssssssssssssss}$\\
\hline \hline
$f_{\beta_1(1c)}(y)$&$-(D-2)(D-1)\;\;I^4 [F C^{(4)}]$\\
\hline \hline
\hline
\multicolumn{2}{|c|}{}\\
\multicolumn{2}{|c|}{${\rm Total\;for\;}f_{\beta_1}(y)$}\\
\multicolumn{2}{|c|}{}\\
\hline \hline
\multicolumn{2}{|c|}{}\\
\multicolumn{2}{|c|}{$\frac{(D-1)(D+2)(D+4) H^{2D-4}\pi^{-D}\Gamma{(\frac{D}2-1})^2}{128(D+1)} \; (\frac1{y})^{D-2} + \frac{9H^4}{16\pi^4} \; \frac1{y}.$}\\
\multicolumn{2}{|c|}{}\\
\end{tabular}\label{tb1}
\end{ruledtabular}
\end{table}

\begin{table}[htbp] 
\caption{For $\beta_2 \equiv  aa' (a^2+a'^2) H^2 \square$ type terms.}
\begin{ruledtabular}
\begin{tabular}{|l|l|}
\hline
\multicolumn{2}{|c|}{}\\
\multicolumn{2}{|c|}{$f_{\beta_2}(y) = C_{1a}\;\;f_{\beta_2(1A)} (y) + C_{1c}\;\;f_{\beta_2(1C)} (y) , $}\\
\multicolumn{2}{|c|}{}\\
\hline \hline
$f_{\beta_2(1a)}(y)$&$\frac12(D-2)^2\;\;I^4 [ F A^{(4)}]\phantom{sssssssssssssssssssssssssssssssssssssssssssss}$\\
\hline \hline
$f_{\beta_2(1c)}(y)$&$\frac12(D-2)^2\;\;I^4 [ F C^{(4)}]$\\
\hline \hline
\hline
\multicolumn{2}{|c|}{}\\
\multicolumn{2}{|c|}{${\rm Total\;for\;}f_{\beta_2}(y)$}\\
\multicolumn{2}{|c|}{}\\
\hline \hline
\multicolumn{2}{|c|}{}\\
\multicolumn{2}{|c|}{$-\frac{(D-2)(D+2)(D+4) H^{2D-4}\pi^{-D}\Gamma{(\frac{D}2-1})^2 }{256(D+1)} \; (\frac1{y})^{D-2} - \frac{3H^4}{16\pi^4}  \; \frac1{y}.$}\\
\multicolumn{2}{|c|}{}\\
\end{tabular}\label{tb2}
\end{ruledtabular}
\end{table}

\begin{table}[htbp] 
\begin{center}
\caption{For $\gamma_1 \equiv (aa')^2 H^4 $ type terms.}
\begin{ruledtabular}
\begin{tabular}{|l|l|}
\hline
\multicolumn{2}{|c|}{}\\
\multicolumn{2}{|c|}{$f_{\gamma_1}(y) =C_{1a}\;\;f_{\gamma_1(1a)} (y)+C_{2a}\;\;f_{\gamma_1(2a)} (y)+C_{3a}\;\;f_{\gamma_1(3a)}(y)+C_{b} \;\;f_{\gamma_1(b)} (y)+C_{1c}\;\;f_{\gamma_1(1c)} (y)+C_{2c}\;\;f_{\gamma_1(2c)} (y)+C_{3c}\;\;f_{\gamma_1(3c)} (y) , $}\\
\multicolumn{2}{|c|}{}\\
\hline \hline
&$\frac12 (D-2)^3 (D-1) I^4[F A^{(4)}]-2(D-1)(I^2[F' A'] + I^3[F' A''] + I^4[F' A^{(3)}])$\\
$f_{\gamma_1(1a)}(y)$&$\;\;+(12+40y-26y^2)(I[F'A''] + I^2[F'A^{(3)}])+8(D-1) (I^2[F'A''] + I^3[F'A^{(3)}])+(-48y+64y^2-20y^3) I[F'A^{(3)}]$\\
&$\;\;+(D-1)(-24 I^2[F'A^{(3)}] + 28 I[y I[F'A^{(3)}]] -14 I[y^2 I[F'A^{(3)}]])-20(D-1) I[y I[F'A''] + y I^2[FA^{(3)}]]$\\
\hline \hline
$f_{\gamma_1(2a)}(y)$&$4(D^2-1) (I[F'A'']-I^2[F'A^{(3)}])$\\
\hline \hline
&$8(D-1)^2 (I^2[F'A''] + I^3[F'A^{(3)}]) -8(D-1) (I[F'A''] + I^2[F' A^{(3)}])+16(D-1) y (I[F'A''] + I^2[F' A^{(3)}])$\\
$f_{\gamma_1(3a)}(y)$&$\;\;+(-16(D-1)y + 8(D-1)y^2) I[F'A^{(3)}] -8(D-1)(D+4) I^2[F'A^{(3)}]$\\
&$\;\;+4(D-1)^2 I[yI[F'A^{(3)}]]$ \\
\hline \hline
&$2(D-1) ( I[F'B'] + I^2[F'B''] + I^3[F'B^{(3)}] )-4(D^2-1) (I^2[F'B''] + I^3[F'B^{(3)}])$\\
$f_{\gamma_1(b)}(y)$&$\;\;+((4-2y)(D-1)) (I[F'B''] +I^2[F'B^{(3)}])+(8(D-1)y - 2(D-1)y^2) I[F'B^{(3)}]$\\
&$\;\;+4((D-1)(D+4)) I^2[F'B^{(3)}] -2((D-1)(D+4)) I[yI[F' B^{(3)}]]$\\
\hline \hline
&$\frac12 (D-2)^3 (D-1) I^4[F C^{(4)}]-2(D-1)(I^2[F'C'] + I^3[F' C''] + I^4[F' C^{(3)}])$\\
$f_{\gamma_1(1c)}(y)$&$\;\;+(12+40y-26y^2)(I[F'C''] + I^2[F'C^{(3)}])+8(D-1) (I^2[F'C''] + I^3[F'C^{(3)}])+(-48y+64y^2-20y^3) I[F'C^{(3)}]$\\
&$\;\;+(D-1)(-24 I^2[F'C^{(3)}] + 28 I[y I[F'C^{(3)}]] -14 I[y^2 I[F'C^{(3)}]])-20(D-1) I[y I[F'C''] + y I^2[FC^{(3)}]]$\\
\hline \hline
$f_{\gamma_1(2c)}(y)$&$ 4(D^2-1)(I[F'C'']-I^2[F'C^{(3)}])$\\
\hline \hline
&$8(D-1)^2 (I^2[F'C''] + I^3[F'C^{(3)}]) -8(D-1) (I[F'C''] + I^2[F' C^{(3)}])+16(D-1) y (I[F'C''] + I^2[F' C^{(3)}])$\\
$f_{\gamma_1(3c)}(y)$&$\;\;+(-16(D-1)y + 8(D-1)y^2) I[F'C^{(3)}] -8(D-1)(D+4) I^2[F'C^{(3)}]$\\
&$\;\;+4(D-1)^2 I[yI[F'C^{(3)}]]$\\
\hline \hline 
\hline
\multicolumn{2}{|c|}{}\\
\multicolumn{2}{|c|}{${\rm Total\;for\;}f_{\gamma_1}(y)$}\\
\multicolumn{2}{|c|}{}\\
\hline \hline
\multicolumn{2}{|c|}{}\\
\multicolumn{2}{|c|}{$\frac{(D-20)(D-2)(D-1)D(D+2)H^{2D-4}\pi^{-D}\Gamma{(\frac{D}2-1})^2}{32(D+1)}\;(\frac{1}{y})^D 
+ \frac{(D-2)(D^5-35 D^4+72 D^3-100 D^2-320 D+256)H^{2D-4}\pi^{-D}\Gamma{(\frac{D}2-1})^2}{256(D+1)}\;(\frac{1}{y})^{D-1}$}\\
\multicolumn{2}{|c|}{}\\
\multicolumn{2}{|c|}{$+\frac{(D^7-51 D^6+240 D^5+532 D^4-5008 D^3+7072 D^2+1472 D-5632) H^{2D-4}\pi^{-D}\Gamma{(\frac{D}2})^2}{1024(D-2)^2(D+1)}\;(\frac{1}{y})^{D-2} -\frac{27H^4}{16\pi^4}\;\frac{1}{y}.$}\\
\multicolumn{2}{|c|}{}\\
\end{tabular}\label{tg1}
\end{ruledtabular}
\end{center}
\end{table}

\begin{table}[htbp] 
\begin{center}
\caption{For $\gamma_2 \equiv aa'(a^2+a'^2) H^4 $ type terms.}
\begin{ruledtabular}
\begin{tabular}{|l|l|}
\hline
\multicolumn{2}{|c|}{}\\
\multicolumn{2}{|c|}{$f_{\gamma_2}(y)=C_{1a}\;\;f_{\gamma_2(1a)} (y)+ C_{3a}\;\;f_{\gamma_2(3a)}(y)+C_{b}\;\;f_{\gamma_2(b)} (y)+C_{1c}\;\;f_{\gamma_2(1c)} (y)+ C_{3c}\;\;f_{\gamma_2(3c)} (y) , $}\\
\multicolumn{2}{|c|}{}\\
\hline \hline
$f_{\gamma_2(1a)}(y)$&$\frac12 (D-2)^2 y I^3[FA^{(4)}]] -32y (I[F'A''] + I^2[F A^{(3)}])- 12(D-1)( I^2[F'A''] + I^3[F'A^{(3)}] )$\\
&$\;\;+ (D-1) (24I^2[F'A^{(3)}]- 12 I[y I[F'A^{(3)}]])+(48y-26y^2) I[F'A^{(3)}]$\\
\hline \hline
$f_{\gamma_2(3a)}(y)$&$4(D-1) ( I[F'A'] + I^2[F'A''] + I^3[F'A^{(3)}])- 12(D-1) (I^2[F'A''] + I^3[F'A^{(3)}])$\\
&$\;\;+8(D-1)y (I[F'A''] + I^2[F'A^{(3)}])+2(D-1)y^2 I[F'A^{(3)}]+24(D-1) I^2[F'A^{(3)}]-12(D-1)I[y I[F' A^{(3)}]]$\\
\hline \hline
$f_{\gamma_2(b)}(y)$&$-12(D-1)  I^2[F'B^{(3)}]$\\
\hline \hline
$f_{\gamma_2(1c)}(y)$&$\frac12 (D-2)^2 y I^3[FB^{(4)}]] -32y (I[F'C''] + I^2[F C^{(3)}])- 12(D-1)( I^2[F'C''] + I^3[F'C^{(3)}] )$\\
&$\;\;+ (D-1) (24I^2[F'C^{(3)}]- 12 I[y I[F'C^{(3)}]])+(48y-26y^2) I[F'C^{(3)}]$\\
\hline \hline
$f_{\gamma_2(3c)}(y)$&$4(D-1) ( I[F'C'] + I^2[F'C''] + I^3[F'C^{(3)}] )- 12(D-1) (I^2[F'C''] + I^3[F'C^{(3)}])$\\
&$\;\;+8(D-1)y (I[F'C''] + I^2[F'C^{(3)}])+2(D-1)y^2 I[F'C^{(3)}]+24(D-1) I^2[F'C^{(3)}]-12(D-1)I[y I[F' C^{(3)}]]$\\
\hline \hline
\hline
\multicolumn{2}{|c|}{}\\
\multicolumn{2}{|c|}{${\rm Total\;for\;}f_{\gamma_2}(y)$}\\
\multicolumn{2}{|c|}{}\\
\hline \hline
\multicolumn{2}{|c|}{}\\
\multicolumn{2}{|c|}{$\frac{3(D-2)D(D^2+D-2)H^{2D-4}\pi^{-D}\Gamma{(\frac{D}2-1})^2}{8(D+1)}\;(\frac{1}{y})^D +\frac{(D-2)(D^4+18 D^3-17 D^2+8 D+8) H^{2D-4}\pi^{-D}\Gamma{(\frac{D}2-1})^2}{64(D+1)} (\frac{1}{y})^{D-1}$}\\ 
\multicolumn{2}{|c|}{}\\
\multicolumn{2}{|c|}{$-\frac{(D^7-55 D^6+536 D^5-2424 D^4+4916 D^3-1840 D^2-5552 D+4544) H^{2D-4}\pi^{-D}\Gamma{(\frac{D}2-1})^2}{1024(D-2)(D+1)}\;(\frac{1}{y})^{D-2} + \frac{3H^4}{16\pi^4}\frac1{y}.$}\\
\multicolumn{2}{|c|}{}\\
\end{tabular}\label{tg2}
\end{ruledtabular}
\end{center}
\end{table}

\begin{table}[htbp]
\begin{center}
\caption{For $\gamma_3 \equiv aa'(a+a)^2 H^4 $ type terms.}
\begin{ruledtabular}
\begin{tabular}{|l|l|}
\hline
\multicolumn{2}{|c|}{}\\
\multicolumn{2}{|c|}{$f_{\gamma_3}(y) = C_{1a}\;\;f_{\gamma_3(1a)} (y)+ C_{1c}\;\;f_{\gamma_3(1c)} (y) , $}\\
\multicolumn{2}{|c|}{}\\
\hline \hline
$f_{\gamma_3(1a)}(y)$&$\frac12 (D-2)^2(y^2I^2[F A^{(4)}] + y I^3[F A^{(4)}]) +\frac12 (D-2)^3 y I^3[F A^{(4)}]\phantom{ssssssssssssssssssssssssssssss}$\\
\hline \hline
$f_{\gamma_3(1c)}(y)$&$\frac12 (D-2)^2(y^2I^2[F C^{(4)}] + y I^3[F C^{(4)}]) +\frac12 (D-2)^3 y I^3[F C^{(4)}]$\\
\hline \hline
\hline
\multicolumn{2}{|c|}{}\\
\multicolumn{2}{|c|}{${\rm Total\;for\;}f_{\gamma_3}(y)$}\\
\multicolumn{2}{|c|}{}\\
\hline \hline
\multicolumn{2}{|c|}{}\\
\multicolumn{2}{|c|}{$\frac{3H^4}{16\pi^4}\;\frac{1}{y}.$}\\
\multicolumn{2}{|c|}{}\\
\end{tabular}\label{tg3}
\end{ruledtabular}
\end{center}
\end{table}

\begin{table}[htbp]
\caption{For $\delta \equiv (a^2+a'^2) H^2 \nabla^2 \square $ type terms.}
\begin{ruledtabular}
\begin{tabular}{|l|l|}
\hline
\multicolumn{2}{|c|}{}\\
\multicolumn{2}{|c|}{$f_{\delta}(y) = C_{1a}\;\;f_{\delta (1a)}(y)+C_{3a}\;\;f_{\delta (3a)}(y)+C_{b}\;\;f_{\delta (b)}(y) +C_{1a}\;\;f_{\delta (1c)}(y)+C_{3c}\;\;f_{\delta (3c)}(y) , $}\\
\multicolumn{2}{|c|}{}\\
\hline \hline
$f_{\delta (1a)}(y)$&$\phantom{ssss}- I^4 [ F A^{(4)}]$\\
\hline \hline
$f_{\delta (3a)}(y)$&$\phantom{ssss}- I^4 [ F A^{(4)}]$\\
\hline \hline
$f_{\delta (b)}(y)$&$\phantom{sssss}\frac12 I^4 [ F B^{(4)}]$\\
\hline \hline
$f_{\delta (1c)}(y)$&$\phantom{ssss}- I^4 [ F C^{(4)}]$\\
\hline \hline
$f_{\delta (3c)}(y)$&$\phantom{ssss}- I^4 [ F C^{(4)}]$\\
\hline \hline
\hline
\multicolumn{2}{|c|}{}\\
\multicolumn{2}{|c|}{${\rm Total\;for\;}f_{\delta}(y)$}\\
\multicolumn{2}{|c|}{}\\
\hline \hline 
\multicolumn{2}{|c|}{}\\
\multicolumn{2}{|c|}{$\frac{H^4}{32\pi^4}\frac1{y}.$}\\
\multicolumn{2}{|c|}{}\\
\end{tabular}\label{td}
\end{ruledtabular}
\end{table}

\begin{table}[htbp] 
\begin{center}
\caption{For $\epsilon_1 \equiv a a' H^2 \nabla^2 $ type terms.}
\begin{ruledtabular}
\begin{tabular}{|l|l|}
\hline
\multicolumn{2}{|c|}{}\\
\multicolumn{2}{|c|}{$f_{\epsilon_1}(y) = C_{1a} \;\;f_{\epsilon_1(1a)} (y)+ C_{2a} \;\;f_{\epsilon_1(2a)} (y)+ C_{3a} \;\;f_{\epsilon_1(3a)}(y) + C_{b}\;\;f_{\epsilon_1(b)} (y) + C_{1c}\;\;f_{\epsilon_1(1C)} (y)+ C_{2c}\;\;f_{\epsilon_1(2c)} (y)+ C_{3c}\;\;f_{\epsilon_1(3c)} (y) , $}\\
\multicolumn{2}{|c|}{}\\
\hline \hline
&$\frac12 (D-2) 2 D  I^4[F A^{(4)}] + 2(D-2) I[y I^2[FA^{(4)}] ] -\frac12 (D-2) I[y^2 I^2[FA^{(4)}] ] -\frac12 (D-2)D I[y I^3[FA^{(4)}]] $\\
$f_{\epsilon_1(1a)}(y)$&$\;-\frac14 (D-2)^3 I^5[FA^{(4)}] -(D-2)^2 I^4[F A^{(4)}] -(D-2)(D-1) I^4[FA^{(4)}] +(I^3[F'A'] + I^4[F'A''] + I^5[F'A^{(3)}])$\\
&$\;+10 I^2[y I[F'A''] + y I^2[F'A^{(3)}] ] -4( I^3[F'A''] + I^4[F'A^{(3)}] ) + 7 I^2 [y^2 I[F'A^{(3)}] ]- 14 I^2 [y I[F'A^{(3)}] ] +12 I^3[F'A^{(3)}]$\\
\hline \hline
$f_{\epsilon_1(2a)}(y)$&$4(D+1)I^3[F'A^{(3)}]$\\
\hline \hline
$f_{\epsilon_1(3a)}(y)$&$-(D-2)(D-1)I^4[FA^{(4)}] + 4(D-1) (I^3[F'A''] +I^4[F'A^{(3)}])+ 4(D+4) I^3[F'A^{(3)}] -2(D-1) I^2 [y I[F'A^{(3)}]]$\\
\hline \hline
$f_{\epsilon_1(b)}(y)$&$\frac12 (D-2)(D-1) I^4 [FB^{(4)}]+ 2(D+1) ( I^3[F'B'']+I^4[F'B^{(3)}] )+ (D+4) I^2[y I[F'B^{(3)}]] - 2(D+4) I^3[F'B ^{(3)}]$\\
\hline \hline
&$\frac12 (D-2) 2 D  I^4[F C^{(4)}] + 2(D-2) I[y I^2[FC^{(4)}] ] -\frac12 (D-2) I[y^2 I^2[FC^{(4)}] ] -\frac12 (D-2)D I[y I^3[FC^{(4)}]] $\\
$f_{\epsilon_1(1c)}(y)$&$\;-\frac14 (D-2)^3 I^5[FC^{(4)}] -(D-2)^2 I^4[F C^{(4)}] -(D-2)(D-1) I^4[FC^{(4)}]+(I^3[F'C'] + I^4[F'C''] + I^5[F'C^{(3)}])$\\
&$ \;+10 I^2[y I[F'C''] + y I^2[F'C^{(3)}] ] -4( I^3[F'C''] + I^4[F'C^{(3)}] ) + 7 I^2 [y^2 I[F'C^{(3)}] ]- 14 I^2 [y I[F'C^{(3)}] ] +12 I^3[F'C^{(3)}]$\\
\hline \hline
$f_{\epsilon_1(2c)}(y)$&$4(D+1)I^3[F'C^{(3)}]$\\
\hline \hline
$f_{\epsilon_1(3c)}(y)$&$-(D-2)(D-1)I^4[FC^{(4)}] + 4(D-1) (I^3[F'C''] +I^4[F'C^{(3)}])+ 4(D+4) I^3[F'C^{(3)}] -2(D-1) I^2 [y I[F'C^{(3)}] ]$\\ 
\hline \hline
\hline
\multicolumn{2}{|c|}{}\\
\multicolumn{2}{|c|}{${\rm Total\;for\;}f_{\epsilon_1}(y)$}\\
\multicolumn{2}{|c|}{}\\
\hline \hline
\multicolumn{2}{|c|}{}\\
\multicolumn{2}{|c|}{$\frac{(D^4+12 D^3-24 D^2-20 D+16) H^{2D-4}\pi^{-D}\Gamma{(\frac{D}2-1})^2}{64(D^2-1)}\;(\frac{1}{y})^{D-2} +\frac{27H^4}{32\pi^4}\;\frac{1}{y}.$}\\
\multicolumn{2}{|c|}{}\\
\end{tabular}\label{te1}
\end{ruledtabular}
\end{center}
\end{table}

\begin{table}[htbp] 
\begin{center}
\caption{For $\epsilon_2 \equiv (a^2 + a^{\prime 2}) H^2 \nabla^2 $ type terms.}
\begin{ruledtabular}
\begin{tabular}{|l|c|}
\hline
\multicolumn{2}{|c|}{}\\
\multicolumn{2}{|c|}{$f_{\epsilon_2}(y) = C_{1a}\;\;f_{\epsilon_2(1a)} (y) + C_{3a}\;\;f_{\epsilon_2(3a)} (y)+ C_{b} \;\;f_{\epsilon_2(b)} (y) + C_{1c} \;\;f_{\epsilon_2(1c)} (y) + C_{3c}\;\;f_{\epsilon_2(3c)} (y) , $}\\
\multicolumn{2}{|c|}{}\\
\hline \hline
$f_{\epsilon_2(1a)}(y)$&$ 6 ( I^3[F'A'']+I^4[F 'A^{(3)}] ) + 6 I^2[y I[F'A^{(3)}]]-12 I^3[F'A^{(3)}] \phantom{ssssssssssssssssssssssssss}$\\
\hline \hline
$f_{\epsilon_2(3a)}(y)$&$ 6 ( I^3[F'A'']+I^4[F 'A^{(3)}] ) + 6 I^2[y I[F'A^{(3)}]]-12 I^3[F'A^{(3)}] \phantom{ssssssssssssssssssssssssss}$\\
\hline \hline
$f_{\epsilon_2(b)}(y)$&$ \frac12 y I^3[FB^{(4)}] + 6 I^3[F'B^{(3)}] \phantom{ssssssssssssssssssssssssss}$\\
\hline \hline
$f_{\epsilon_2(1c)}(y)$&$ 6 ( I^3[F'C'']+I^4[F 'C^{(3)}] ) + 6 I^2[y I[F'C^{(3)}]]-12 I^3[F'C^{(3)}] \phantom{ssssssssssssssssssssssssss}$\\
\hline \hline
$f_{\epsilon_2(3c)}(y)$&$ 6 ( I^3[F'C'']+I^4[F 'C^{(3)}] ) + 6 I^2[y I[F'C^{(3)}]]-12 I^3[F'C^{(3)}] \phantom{ssssssssssssssssssssssssss}$\\
\hline \hline
\hline
\multicolumn{2}{|c|}{}\\
\multicolumn{2}{|c|}{${\rm Total\;for\;}f_{\epsilon_2}(y)$}\\
\multicolumn{2}{|c|}{}\\
\hline \hline
\multicolumn{2}{|c|}{}\\
\multicolumn{2}{|c|}{$-\frac{(D-2)(13D^2+6D-16) H^{2D-4}\pi^{-D}\Gamma{(\frac{D}2-1})^2}{128(D^2-1)}\;(\frac{1}{y})^{D-2}.$}\\
\multicolumn{2}{|c|}{}\\
\end{tabular}\label{te2}
\end{ruledtabular}
\end{center}
\end{table}

\begin{table}[htbp] 
\begin{center}
\caption{For $\epsilon_3 \equiv (a + a')^2 H^2 \nabla^2 $ type terms.}
\begin{ruledtabular}
\begin{tabular}{|l|c|}
\hline
\multicolumn{2}{|c|}{}\\
\multicolumn{2}{|c|}{$f_{\epsilon_3}(y) = C_{1a}\;\;f_{\epsilon_3(1a)} (y)+  C_{3a} \;\;f_{\epsilon_3(3a)} (y)+ C_{b} \;\;f_{\epsilon_3(b)} (y) + C_{1c}\;\;f_{\epsilon_3(1c)} (y)+C_{3c} \;\;f_{\epsilon_3(3c)} (y),$}\\
\multicolumn{2}{|c|}{}\\
\hline \hline
$f_{\epsilon_3(1a)}(y)$&$ -(D-2)y I^3[FA^{(4)}] \phantom{ssssssssssssssssssssssssssssssssssssssssss}$\\
\hline \hline
$f_{\epsilon_3(3a)}(y)$&$ -(D-2)y I^3[FA^{(4)}] \phantom{ssssssssssssssssssssssssssssssssssssssssss}$\\
\hline \hline
$f_{\epsilon_3(b)}(y)$&$ \frac12 y^2 I^2[FB^{(4)}]+ \frac12 (D-1) y I^3[FB^{(4)}] \phantom{ssssssssssssssssssssssssssssssssssssssss}$\\
\hline \hline
$f_{\epsilon_3(1c)}(y)$&$ -(D-2)y I^3[FC^{(4)}] \phantom{ssssssssssssssssssssssssssssssssssssssssss}$\\
\hline \hline
$f_{\epsilon_3(3c)}(y)$&$ -(D-2)y I^3[FC^{(4)}] \phantom{ssssssssssssssssssssssssssssssssssssssssss}$\\
\hline \hline
\hline
\multicolumn{2}{|c|}{}\\
\multicolumn{2}{|c|}{${\rm Total\;for\;}f_{\epsilon_3}(y)$}\\
\multicolumn{2}{|c|}{}\\
\hline \hline
\multicolumn{2}{|c|}{}\\
\multicolumn{2}{|c|}{$\frac{(D+2)(D+4)H^{2D-4}\pi^{-D}\Gamma{(\frac{D}2})^2}{32(D^2-1)}\;(\frac{1}{y})^{D-2}-\frac{H^4}{16\pi^4}\;\frac{1}{y}.$}\\
\multicolumn{2}{|c|}{}\\
\end{tabular}\label{te3}
\end{ruledtabular}
\end{center}
\end{table}

\begin{table}[htbp] 
\begin{center}
\caption{For $\zeta \equiv \nabla^4 $ type terms.}
\begin{ruledtabular}
\begin{tabular}{|l|c|}
\hline
\multicolumn{2}{|c|}{}\\
\multicolumn{2}{|c|}{$f_{\zeta}(y) = C_{1a}\;\;f_{\zeta (1a)} (y)+C_{2a}\;\;f_{\zeta (2a)} (y)+ C_{3a}\;\;f_{\zeta (3a)} (y) + C_{b}\;\;f_{\zeta (b)} (y)+C_{1c}\;\;f_{\zeta (1c)} (y) +C_{2c}\;\;f_{\zeta (2c)} (y)+C_{3c}\;\;f_{\zeta (3c)} (y),$}\\
\multicolumn{2}{|c|}{}\\
\hline \hline
$f_{\zeta (1a)}(y)$&$ I^4[F A^{(4)}] +\frac12 (D-2) I^5[F A^{(4)}]) \phantom{ssssssssssssssss}$\\
\hline \hline
$f_{\zeta (2a)}(y)$&$ I^4[F A^{(4)}] \phantom{ssssssssssssssss}$\\
\hline \hline
$f_{\zeta (3a)}(y)$&$  2 I^4[FA^{(4)}] +\frac12 (D-2) I^5[F A^{(4)}] \phantom{ssssssssssssssss}$\\
\hline \hline
$f_{\zeta (b)}(y)$&$ -I^4[FB^{(4)}] -\frac14 (D-2) I^5[F B^{(4)}] \phantom{ssssssssssssssss}$\\
\hline \hline
$f_{\zeta (1c)}(y)$&$ I^4[F C^{(4)}] +\frac12 (D-2) I^5[F C^{(4)}]) \phantom{ssssssssssssssss}$\\
\hline \hline
$f_{\zeta (2c)}(y)$&$ I^4[F C^{(4)}] \phantom{ssssssssssssssss}$\\
\hline \hline
$f_{\zeta (3c)}(y)$&$  2 I^4[FC^{(4)}] +\frac12 (D-2) I^5[F C^{(4)}] \phantom{ssssssssssssssss}$\\
\hline \hline
\hline
\multicolumn{2}{|c|}{}\\
\multicolumn{2}{|c|}{${\rm Total\;for\;}f_{\zeta}(y)$}\\
\multicolumn{2}{|c|}{}\\
\hline \hline
\multicolumn{2}{|c|}{}\\
\multicolumn{2}{|c|}{$0.$}\\
\multicolumn{2}{|c|}{}\\
\end{tabular}\label{tz}
\end{ruledtabular}
\end{center}
\end{table}


\begin{thebibliography}{}

\providecommand{\url}[1]{\textit{#1}}
\providecommand{\urlprefix}{}
\providecommand{\eprint}[2][]{\url{#2}}


\bibitem{Schr} E. Schr\"odinger, Physica {\bf 6} (1939) 899.

\bibitem{Parker1} L. Parker, Phys. Rev. Lett. {\bf 21} (1968) 562;
Phys. Rev. {\bf 183} (1969) 1057;
Phys. Rev. {\bf D3} (1971) 346.

\bibitem{Parker2} L. H. Ford and L. Parker, Phys. Rev. D {\bf 16} (1977) 1601.

\bibitem{SMC} A. A. Starobinsky, JET Lett. {\bf 30} (1979) 682;
V. F. Mukhanov and G. V. Chibisov, JETP Lett. {\bf 33} (1981) 532.

\bibitem{WMAP9}
  G.~Hinshaw {\it et al.}  [WMAP Collaboration],
  Astrophys.\ J.\ Suppl.\  {\bf 208} (2013) 19,
  arXiv:1212.5226.

\bibitem{Planck16} 
  P.~A.~R.~Ade {\it et al.}  [Planck Collaboration],
  arXiv:1303.5076. 

\bibitem{BICEP2} 
  P.~A.~R.~Ade {\it et al.}  [BICEP2 Collaboration],
  Phys.\ Rev.\ Lett.\  {\bf 112}, 241101 (2014),
  arXiv:1403.3985.

\bibitem{phi4} T. Brunier, V. K. Onemli and R. P. Woodard, Class. Quant.
Grav. {\bf 22} (2005) 59, gr-qc/0408080; E. O. Kahya and V. K.
Onemli, Phys. Rev. D {\bf 76} (2007) 043512, gr-qc/0612026.

\bibitem{SQED} T. Prokopec, O. T\"ornkvist and R. P. Woodard, Phys. Rev.
Lett. {\bf 89} (2002) 101301, astro-ph/0205331; Annals Phys. {\bf
303} (2003) 251, gr-qc/0205130; T. Prokopec and R. P. Woodard, Am.
J. Phys. {\bf 72} (2004) 62, astro-ph/0303358; Annals Phys. {\bf
312} (2004) 1, gr-qc/0310056.

\bibitem{PP} T. Prokopec and E. Puchwein, JCAP {\bf 0404} (2004) 007, astro-ph/0312274.

\bibitem{Yukawa} T. Prokopec and R. P. Woodard, JHEP {\bf 0310} (2003)
059, astro-ph/0309593; 
B. Garbrecht and T. Prokopec, Phys. Rev. D {\bf 73} (2006) 064036, gr-qc/0602011; L. D. Duffy and R. P. Woodard,
Phys. Rev. D {\bf 72} 024023, hep-ph/0505156.

\bibitem{mw} S. P. Miao and R. P. Woodard, Class. Quant. Grav. {\bf 23}
(2006) 1721, gr-qc/0511140; Phys. Rev. D {\bf 74} (2006) 024021,
gr-qc/0603135; Class. Quant. Grav. {\bf 25} (2008) 145009,
arXiv:0803.2377.

\bibitem{SPM} S. P. Miao, arXiv:0705.0767; Phys. Rev. D {\bf 86} (2012)
104051, arXiv:1207.5241.

\bibitem{KW1} E. O. Kahya and R. P. Woodard, Phys. Rev. D {\bf 72} (2005)
104001, gr-qc/0508015; Phys. Rev. D {\bf 74} (2006) 084012, gr-qc/0608049.

\bibitem{kahya} E. O. Kahya and R. P. Woodard, Phys. Rev. D \textbf{76}, 124005 (2007), arXiv:0709.0536.

\bibitem{kahya2} E. O. Kahya and R. P. Woodard, Phys. Rev. D {\bf 77} (2008) 084012, arXiv:0710.5282.

\bibitem{PW} S. Park and R. P. Woodard, 
Phys. Rev. D {\bf 83} (2011) 084049, arXiv:1101.5804; Phys. Rev. D {\bf 84} (2011)
124058, arXiv:1109.4187.

\bibitem{LW} K. E. Leonard and R. P. Woodard, 
Class. Quant. Grav. {\bf 31} (2014) 015010, arXiv:1304.7265.

\bibitem{PMTW} P. J. Mora, N. C. Tsamis and R. P. Woodard, JCAP
{\bf 10} (2013) 018, arXiv:1307.1422.

\bibitem{LPPW} K. E. Leonard, S. Park, T. Prokopec and R. P. Woodard, 
Phys. Rev. D \textbf{90} (2014) 024032, arXiv:1403.0896

\bibitem{CV} A. Campos and E. Verdaguer, Phys. Rev. D \textbf{49} (1994) 1861; Phys. Rev. D \textbf{53} (1996) 1927.

\bibitem{FRV1} M. B. Fr\"ob, A. Roura and E. Verdaguer, JCAP \textbf{1208} (2012) 009, arXiv:1205.3097.

\bibitem{FPRV} M. B. Fr\"ob, D. B. Papadopoulos, A. Roura, E. Verdaguer, Phys. Rev. D. \textbf{87} (2013) 064019, arXiv:1391.5261.
 
\bibitem{FRV2} M. B. Fr\"ob, A. Roura and E. Verdaguer, JCAP \textbf{1407} (2014) 048, arXiv:1403.3335.

\bibitem{Frob} M. B. Fr\"ob, arXiv:1409.7964.
\bibitem{DS2}  D. Seery, JCAP {\bf 0711} (2007) 025, arXiv:0707.3377.

\bibitem{DS1}  D. Seery, JCAP {\bf 0802} (2008) 006, arXiv:0707.3378.

\bibitem{UM} Y. Urakawa and K. Maeda, Phys. Rev. D {\bf 78} 064004 (2008), arXiv:0801.0126.

\bibitem{DS3}  D. Seery, Class. Quantum Grav.  {\bf 27} (2010) 124005, arXiv:1005.1649.
\bibitem{MM} D. Marolf and I. A. Morrison, Phys. Rev.\textbf{D} 82 (2010) 105032, arXiv:1006.0035.
\bibitem{ABG}  V. Assassi, D. Baumann and Daniel Green, JHEP {\bf 1302} (2013) 151, arXiv:1210.7792.

\bibitem{YK} A. Youssef and D. Kreimer, Phys. Rev. D {\bf 89}, 124021 (2014), arXiv:1301.3205.

\bibitem{APS}  E. T. Akhmedov, F. K. Popov and V. M. Slepukhin, Phys. Rev. D {\bf 88}, 024021 (2013), arXiv:1303.1068.

\bibitem{AK} A. Kaya, Phys. Rev. D {\bf 90}, 043506 (2014), arXiv:1306.3236.

\bibitem{YU} Y. Urakawa and K. Maeda, Class.Quant.Grav. {\bf 30} (2013) 233001, arXiv:1306.4461.

\bibitem{LBH} L. Lello, D. Boyanovsky and Richard Holman, Phys. Rev. D {\bf 89}, 063533 (2014), arXiv:1307.4066.

\bibitem{Akh} E.T. Akhmedov, Int. Jour. of Mod. Phys. D {\bf 23}, (2014) 1430001, arXiv:1309.2557.

\bibitem{GRZ} B. Garbrecht, G. Rigopoulos and Y. Zhu , Phys. Rev. D {\bf 89}, 063506 (2014), arXiv:1310.0367.
\bibitem{MP} S. P. Miao and S. Park, 
Phys. Rev. D {\bf 89} (2014) 064053, arXiv:1306.4126.

\bibitem{BD} N. D. Birrell and P. C. W. Davies, {\it Quantum Fields in Curved
Space} (Cambridge University Press, 1982).

\bibitem{tw} N. C. Tsamis and R. P. Woodard, Commun. Math. Phys. \textbf{162}, 217 (1994).

\bibitem{ow1} V. K. Onemli and R. P. Woodard, Classical Quantum Gravity \textbf{19}, 4607 (2002).

\bibitem{ow2} V. K. Onemli and R. P. Woodard, Phys. Rev. D \textbf{70}, 107301 (2004).

\bibitem{af} B. Allen and A. Folacci, Phys. Rev. D \textbf{35}, 3771 (1987).

\bibitem{bogo} N. N. Bogoliubov and O. Parasiuk, Acta Math. \textbf{97}, 227, (1957)
               ; K. Hepp, Commun. Math. Phys. \textbf{2}, 301, (1966)
               ; W. Zimmermann, Commun. Math. Phys. \textbf{11}, 1, (1968)
               ; \textbf{15}, 208, (1969); \textit{Lectures on Elementary Particles and 
               Quantum Field Theory}, edited by S. Deser, M. Grisaru, and H. Pendleton (MIT Press, Cambridge, 1971), Vol. I. 
\bibitem{Sch} J. Schwinger, J. Math. Phys. {\bf 2} 407 (1961).

\bibitem{KTM} K. T. Mahanthappa, Phys. Rev. {\bf 126} 329 (1962) .

\bibitem{LVK} L. V. Keldysh, Sov. Phys. JETP {\bf 20} 1018 (1965).

\bibitem{Jrd} R. D. Jordan, Phys. Rev. D {\bf 33}, 444 (1986).

\bibitem{CHU} E. Calzetta and B. L. Hu, Phys. Rev. D {\bf 35}, 495 (1987).

\bibitem{CVE} A. Campos and E. Verdaguer, Phys. Rev. D {\bf 49}, 1861 (1994).

\bibitem{FW} L. Ford and R. P. Woodard, Clas. Qua. Grav. {\bf 22}, 1637 (2005).

\bibitem{SW11} S. Weinberg, Phys. Rev. D {\bf 72}, 043514 (2005). 
\bibitem{W-QGage} R. P. Woodard, arXiv:1407.4748.


\end{thebibliography}
\end{document}